\documentclass[12pt]{article}
\pdfoutput=1
\usepackage{graphicx}
\usepackage{epstopdf}
\usepackage{amsmath}
\usepackage{amsfonts}
\usepackage{amssymb}
\usepackage{color}
\usepackage{mathrsfs}

\usepackage{graphicx} \usepackage{bm} \usepackage{epsfig}
\usepackage{amsmath, amssymb}
\usepackage{hyperref}
\usepackage{setspace}
\usepackage{color}
\usepackage{colordvi}
\usepackage{comment}
\usepackage{graphicx}

\newcommand{\be}{\begin{equation}}
\newcommand{\ee}{\end{equation}}
\newcommand{\bea}{\begin{eqnarray}}
\newcommand{\eea}{\end{eqnarray}}
\newcommand{\barr}{\begin{array}}
\newcommand{\earr}{\end{array}}
\newcommand{\dl}{\delta}

\newcommand{\eeqref}{\eqref}
\newcommand{\toUV}{\substack{\longrightarrow \\ {\tiny UV}}}
\newcommand{\toepsilon}{\substack{\longrightarrow \\ {\tiny \epsilon\to 0}}}

\def\beq{\begin{equation}}
\def\eeq{\end{equation}}
\def\be{\begin{equation}}
\def\ee{\end{equation}}
\def\bea{\begin{eqnarray}}
\def\eea{\end{eqnarray}}
\def\d{{\partial}}

\def\nn{\nonumber}

\newcommand{\pd}{\partial}
\newcommand{\bit}{\begin{itemize}}
\newcommand{\eit}{\end{itemize}}

\def\calo{{\cal O}}

\newcommand{\bx}{{\bf x}}

\newcommand{\beqa}
{\begin{eqnarray}}
\newcommand{\eeqa}{\end{eqnarray}}

\def\cH{\mathcal{H}}

\def\fun#1#2{\lower3.6pt\vbox{\baselineskip0pt\lineskip.9pt
        \ialign{$\mathsurround=0pt#1\hfill##\hfil$\crcr#2\crcr\sim\crcr}}}

\begin{document}

\begin{center}

{\Large \bf The Lagrangian-space Effective Field Theory\\[0.3cm] of Large Scale Structures}
\\[0.7cm]
{\large Rafael A. Porto${}^{1,2}$, Leonardo Senatore${}^{3,4,5}$  and Matias Zaldarriaga${}^1$}
\\[0.7cm]

{\normalsize { \sl $^{1}$ School of Natural Sciences, Institute for Advanced Study, \\Olden Lane, 
Princeton, NJ 08540, USA}}\\
\vspace{.3cm}

{\normalsize { \sl $^{2}$ Deutsches Elektronen-Synchrotron DESY, Theory Group, D-22603 Hamburg, Germany}}\\
\vspace{.3cm}

{\normalsize { \sl $^{3}$ Stanford Institute for Theoretical Physics, \\ Stanford University, Stanford, CA 94306}}\\
\vspace{.3cm}

{\normalsize { \sl $^{4}$ Kavli Institute for Particle Astrophysics and Cosmology, \\ Stanford University and SLAC, Menlo Park, CA 94025}}\\
\vspace{.3cm}

{\normalsize {\sl $^{\rm 5}$ CERN, Theory Division, 1211 Geneva 23, Switzerland}}\\
\vspace{.3cm}

\vspace{.3cm}

\end{center}

\vspace{.8cm}

\hrule \vspace{0.3cm}
{\small  \noindent \textbf{Abstract} \\[0.3cm]
\noindent We introduce a Lagrangian-space Effective Field Theory (LEFT) formalism for the study of cosmological large scale structures. Unlike the previous Eulerian-space construction, it is naturally formulated as an effective field theory of extended objects in Lagrangian space. In LEFT the resulting finite size effects are described using a multipole expansion parameterized by a set of time dependent coefficients and organized in powers of the ratio of the wavenumber of interest $k$ over the non-linear scale $k_{\rm NL}$. The multipoles encode the effects of the short distance modes on the long-wavelength universe and absorb UV divergences when present. There are no IR divergences in LEFT.  Some of the parameters that control the perturbative approach are not assumed to be small and can be automatically resummed. We present an illustrative one-loop calculation for a power law universe. We describe the dynamics both at the level of the equations of motion and through an action formalism. 
 \vspace{0.3cm}
\hrule
\newpage
\tableofcontents
\newpage
\section{Introduction}

The discovery of the acceleration of the expansion of the universe dramatically changed our picture of Cosmology and has motivated precision studies of the expansion history and growth of structure in an attempt to gather more evidence.  The Baryonic Acoustic Oscillation (BAO) technique and the study of weak lensing by Large Scale Structure (LSS) have emerged as very powerful techniques to constrain the properties of the ``dark-energy".  Extremely ambitious observational programs are now under way to make very precise measurements of the LSS with the goal of making sub-percent measurements of the properties of the dark energy. These surveys will map vast volumes to measure the required number of modes to overcome the statistical noise intrinsic in the comparison between theoretical predictions and data, the cosmic variance.

These same modes can also be used to infer properties of the initial seeds of structure and thus constrain the physics of the early universe, when the initial fluctuations were generated. In particular after the recent results form the Planck satellite, improvements in constraints on non-Gaussianity will have to come from the study of LSS. The combination of  vast amounts of new data and the interesting theoretical problems that these data can address motivates new efforts to make precise theoretical predictions for LSS.

In many respects the tool of choice to study LSS theoretically are numerical simulations. For example, understanding the LSS produced in a universe with only cold dark matter can be considered a solved problem. At least in the sense that numerical simulations can in principle be run with exquisite understanding and control of the numerics and the results used to ``observe" any statistic of choice and thus compute its theoretically predicted value. 
But even without considering the physical processes related to baryons that make first principle numerical simulations currently impossible, even for dark matter only the situation is not fully satisfactory if one does not have a good analytical understanding. An example of how analytical understanding can lead to practical improvement  is the reconstruction technique for the BAO~\cite{Eisenstein:2006nk}. In that case an understanding of the dynamics based on perturbations theory can be used to develop a measurement techniques that sharpens the BAO feature in the correlation function undoing at least partially the degradation caused by the non-linear dynamics. In a sense non-linearities  moved information from the two point function to higher order correlations. Thus by combining those higher order moments judiciously one can tighten cosmological constraints. Of course one tests these ideas using simulations, but it is only through the analytical understanding of the dynamics that one can propose the new techniques. 

As data improves however the analytical understanding required to develop these type of improved methods will be more stringent. The non-linear effects on the BAO scales are rather small and dominated by large scale motions  produced by relatively linear modes so BAO reconstruction is perhaps not so demanding on our analytical understanding (although of course the fact that one is dealing with biased tracers complicates matters). But as we strive to model modes closer to the non-linear regime, for example to improve constraints on non-Gaussanity, our analytical techniques will have to pass extremely stringent tests. 


Perturbation theory for LSS has a long and distinguished history dating back to the very early days of modern Cosmology {\it e.g.} \cite{Zeldovich:1969sb,Peebles:LSS}. It is extremely successful at calculating correlators at the lowest order or tree level (for a complete review of perturbation theory results see \cite{Bernardeau:2001qr}.) However results for the first nontrivial correction to tree level results, the ``loop corrections", have been less than satisfactory. These corrections are relevant for upcoming observations and they are not under theoretical control. One example of the failure of loop calculations can be found when considering the simple case of Einstein-de Sitter universes with power law initial conditions. Already the one-loop correction to the power spectrum diverges if the spectral index is not in the range $-3<n<-1$. Simulations with $n>-1$ can be done and the results are of course perfectly well defined. Thus in these cases SPT is making an infinite mistake. There are no divergencies for LCDM cosmology so the field has, with few exceptions~\cite{Scoccimarro:1996se}, ignored the issue. But these divergencies are simply the ``canary in the coal mine" pointing out that there is something fundamentally flawed in the standard approach. This flaw leads to infinite mistakes in some cosmologies and finite mistakes in LCDM. 

Nevertheless even if the errors are finite we need to learn how to track them and estimate their sizes. The reason for the failure is clear. Perturbation theory cannot be used to describe the small scales. The series simply does not converge so no resummation of diagrams will fix the problem. In loop calculations, those small scales affect large scale observables as the loop integral cover all momenta. Thus the errors in the small scales pollute large scale results.

\begin{figure}[h!]
\includegraphics[width=0.8\textwidth]{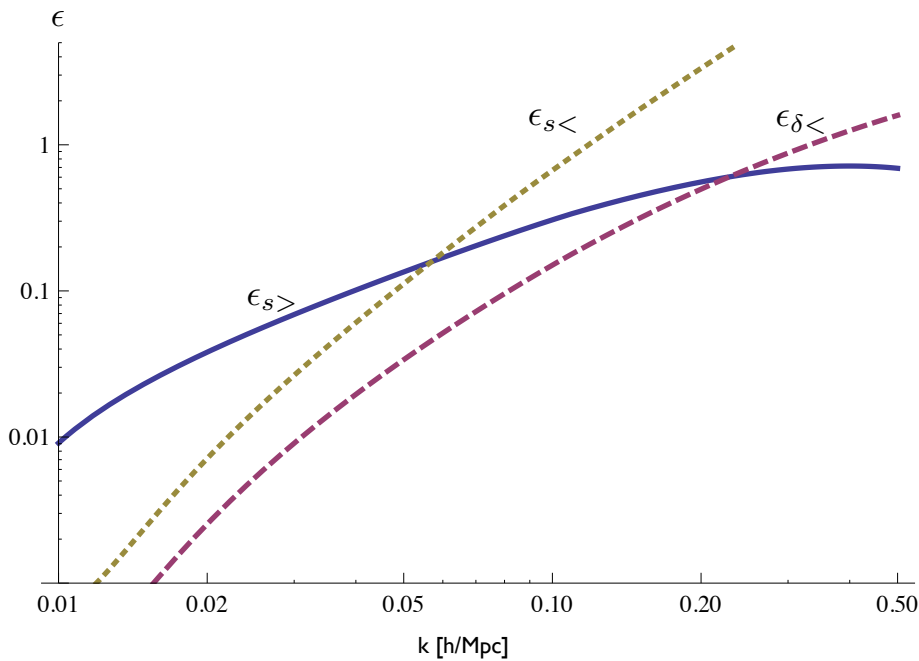}
\caption{\small  Parameters measuring the amplitude of non-linear correction on a mode of wavenumber $k$. They quantify the motions created by modes longer ($\epsilon_{s<}$) and shorter ($\epsilon_{s>}$) than $k$ and the tides from larger scales ($\epsilon_{\delta <}$). }
\label{epsilons}
\end{figure}


This has led to the development of the Effective Theory of Large Scale Structure \cite{Baumann:2010tm,Carrasco:2012cv}. This framework explicitly keeps track of the effects of the small scales using a generalized fluid-like description, where the uncertainties produced by the short distance dynamics are encoded  in a set of coefficients which, from the point of view of perturbation theory, are free parameters to be fitted to either simulations or observations. The fact that the Effective Field Theory (EFT) has free parameters is a virtue rather than a problem, it is for example these free coefficients that allow us to get finite results in cases where Standard Perturbation Theory (SPT) diverges using a standard renormalization procedure. But regardless of this, it allows us to consistently and systematically keep track of the uncertainties produced by the small scales dynamics that lies outside of the regime of applicability of perturbation theory.~\footnote{For other approaches using smoothing, see for example~\cite{Buchert:2005xj,Pueblas:2008uv,Pietroni:2011iz}.}. Until this paper, the EFT has been developed in Eulerian space and used to compute one and two-loop corrections to the matter power spectrum~\cite{Carrasco:2012cv,Carrasco:2013mua} and to study the divergencies that appear in power law universes~\cite{Pajer:2013jj,Carrasco:2013sva}~\footnote{In spite of the problems with SPT discussed in this paper, the computation of higher order loop effects remains an important task to improve existing predictions after adopting a controlled and systematic framework as the one we introduce. For the {state of the art} three-loop results in SPT see \cite{3loop}.}.


Irrespective of these development, in the last few years it has become apparent that for LCDM perturbation theory in Lagrangian space is significantly better than its Eulerian counterpart. This is particularly evident when studying the non-linear effects in the BAO. Extremely impressive results have been obtained using Lagrangian Perturbation Theory (LPT) both in real and redshift space and also for halos {\it eg.} \cite{Carlson:2012bu}. Furthermore even around the non-linear scale the cross correlation coefficient between the results of an N-body simulation and those of perturbation theory are remarkably better when doing LPT \cite{Tassev:2011ac}.  This motivated us to write the EFT in Lagrangian space, which we will do in this paper. In fact \cite{Tassev:2012cq} studied the relation between particle trajectories in simulations and those computed in LPT and found very high correlation coefficients but also non-trivial transfer functions, pointing to the fact that LPT should be improved.  


The difference between Lagrangian and Eulerian perturbation theory can be traced to the fact that there are several different parameters that control the size of non-linearities. In LCDM cosmologies, which have a nontrivial transfer function, these various effects have very different sizes. Thus it is not fully satisfactory to organize perturbation theory in powers of the power spectrum.  For example, simple inspection  shows that the corrections to the power spectrum at a scale $k$ produced by other modes of wavenumber $q$ depends on several different parameters. They depend on the variance of the density fluctuations produced by modes with $q<k$ ($\epsilon_{\delta <}$) and depend on the displacements produced by modes with $q>k$ ($\delta s_>$) through $\epsilon_{s>}= (k \delta s_>)^2$ ~\footnote{We have defined $\epsilon_{X>} = \int_k^\infty d^3 k /(2\pi)^3 P_X(k)$ and $\epsilon_{X<} = \int_0^k d^3 k /(2\pi)^3 P_X(k)$ where $X=(\delta, s)$ stands for either the density or the displacement, and $P_X(k)$ for its power spectrum.}. The fact that modes larger and smaller than $k$ affect the power spectrum through different parameters is what allows SPT to be non-divergent for equal time correlators power law universes in the range $-3<n<-1$. In this range both  $\epsilon_{\delta <}$ and $\epsilon_{s>}$ are finite. Of course the fact that the result is finite does not guarantee that it is converging to the correct result~\footnote{The fact that perturbation theory does not have ultraviolet divergencies does not mean that it converges to the correct answer. For example S.~Weinberg explains that renormalization is necessary irrespectively of the presence of ultraviolet divergencies as early as page xxii of the preface in his Quantum Field Theory textbook~\cite{Weinberg:1995mt}.}.
 
The displacements produced by modes with $q<k$ ($\delta s_<$) do not affect the small scale dynamics directly but they change the final location of those small scale modes  and thus can significantly affect some statistics through the parameter $\epsilon_{s<}= (k \delta s_<)^2$. In fact $\epsilon_{s<}$ is responsible  the broadening of the acoustic peak that degrades the BAO technique. Figure \ref{epsilons} shows the sizes of these $\epsilon$-parameters.  It is clear that  to achieve a desired accuracy one needs to keep more orders in some of these parameters than in others. The biggest of the parameters are those related to displacements which are dominated by large scale modes and thus are very amenable to perturbation theory. 
 
LPT does not expand in  $\epsilon_{s<}$ which in our universe controls the largest non-linearity in the range of scales of interest for the BAO. Thus it would clearly be advantageous to develop the necessary EFT directly in Lagrangian space, as we do here. 
The Lagrangian-space EFT (LEFT) we will develop is a theory of extended objects, a theory for regions of Lagrangian space of size comparable to the non-linear scale. As a result these regions can deform and have a quadrupole and higher multipole moments which modify the way they gravitate and move in an external potential.  

In LEFT we will not expand in $\epsilon_{s<}$, and therefore our calculations will be performed in an expansion in $\epsilon_{s>}$ and $\epsilon_{\delta<}$, which as can be seen from Figure~\ref{epsilons} grow with $k/k_{\rm NL}$ to some power, with $k$ being the wavenumber of interest and $k_{\rm NL}$ being the wavenumber associated to the non-linear scale. As we will show in great detail in the bulk of this paper, at a given order in $k/k_{\rm NL}$, it is sufficient to consider the mass, quadrupole, octupole, etc., and how these are affected by long wavelength perturbations at linear, quadratic, and higher orders, and so on and so forth. Each of these terms have a well defined power counting in $k/k_{\rm NL}$. This means that, as we make the calculations more and more accurate, new parameters characterizing the deformation of finite-sized objects need to be included. The role of these parameters will be both to correctly encode how these multipole moments respond to external gravitational fields, but also to correct for the mistakes  in the loop integrals which include wavenumbers above the non-linear scale. Implementing this procedure goes under the name of `renormalization'. The EFT power counting rules, that we have just outlined here and we will describe in detail in the bulk of the paper, will determine how many of these parameters need to be kept to achieve a desired accuracy. In fact in the Eulerian EFT for our cosmology, a three-loop calculation is required before needing to introduce more than one EFT parameter in the calculation of the matter power spectrum~\cite{Carrasco:2013mua}. As we will describe, the situation is similar in LEFT, although some subtleties will arise when the displacements due to long-wavelength modes are resummed.


This paper is organized as follows. In sec. \ref{sec:exp} we introduce the effective description of long-wavelength modes in Lagrangian space as a multipole expansion at the level of the equations of motion and explain the power counting rules of LEFT. In sec. \ref{sec:oneloop} we perform an illustrative one-loop computation for a power law universe. In sec. \ref{resum} we discuss the resummation of various terms in LEFT through exponentiation. Finally in sec. \ref{sec:action} we re-derive the dynamics in LEFT via an action formalism. Some details of the computations and further comments and examples are relegated to appendices.

\section{Effective description of long-wavelength modes in Lagrangian space\label{sec:exp}}

In this section we introduce the Lagrangian-space Effective Field Theory for Large Scale Structures. We first summarize the well known equations of motion in Lagrangian space, as a reminder to the reader and also to set up our notation, before we proceed to define the basic objects in LEFT.

\subsection{Lagrangian space dynamics}

Let us start describing the dynamics of dark matter particles interacting through a Newtonian potential in Lagrangian space. We will denote each particle's position as ${\vec z}({\vec q},t)$, with ${\vec q}$ the Lagrangian coordinates such that ${\vec z}({\vec q},0) = {\vec q}$. The variable $\vec q$ is just a label for the dark matter particle which, for simplicity, we have taken to be continuous. The equations we need to solve are:
\bea
\frac{d^2 {\vec z}({\vec q},t)}{dt^2} &=& - \nabla \phi[{\vec z}({\vec q},t)]\ ,\\
\nabla^2\phi(x) &=& 4\pi G \rho(x)\ ,
\eea
where $\phi(z)$ is the gravitational potential. In an expanding universe it is useful to switch to co-moving coordinates, namely $d{\vec r} = a(t) d{\vec z}$,  $d\eta = dt/a(t)$. Then we have
\bea
\Phi &=& \phi + \tfrac{1}{2} \dot {\cal H} {\vec z}^2\ ,\\
\vec u &=&  \vec v - {\cal H} {\vec z}\ ,\\
\rho &=& \bar \rho_m ( 1+ \delta)\ ,
\eea
where ${\cal H} = a H$, $\vec u$ is the peculiar velocity and $\bar\rho_m$ is the mean matter density given by: 
\be
\label{rhoq0}
\bar\rho_m(\eta)=\frac{3}{8\pi G}  { H}^2 \Omega_m\ .
\ee
The resulting dynamics becomes
\bea
\frac{d {\vec u}({\vec q},\eta)}{d \eta} +{\cal H}{\vec u}({\vec q},\eta) &=&  - \vec\partial_x \Phi[\vec z({\vec q},\eta)]\ ,\\
\partial_x^2\Phi({\vec x},\eta) &=& \frac{3}{2} {\cal H}^2 \Omega_m \delta({\vec x},\eta)\ . 
\eea
Denoting the displacement as 
\beq {\vec z} = {\vec q} + {\vec s}({\vec q},\eta)\ ,
\eeq 
and using ${\vec u} =  \tfrac{d}{d\eta}{\vec s}$ we can write:
\beq
\label{boxed0}
\frac{d^2 {\vec z}({\vec q},\eta)}{d \eta^2} +{\cal H}\frac{d{\vec z}({\vec q},\eta)}{d\eta} =  - \vec \partial_x \Phi[{\vec z}({\vec q},\eta)]\ .
\eeq
The standard map between between displacement and density is given by:
\beqa
\label{mapqx}
1+\delta({\vec x},\eta)&=& \int  d^3{\vec q}~\delta^3( {\vec x} - {\vec z}({\vec q},\eta)) \nonumber \\
&=& \left[{\rm det} \left(\tfrac{\partial z^i}{\partial q^j}\right)\right]^{-1} =\left[{\rm det} \left(1+\tfrac{\partial s^i}{\partial q^j}\right)\right]^{-1},
\eeqa
where the second line is evaluated at the solution of ${\vec z}({\vec q},\eta)=\vec x$. For simplicity we assumed this equation has only one solution, otherwise the second line involves a sum over all solutions. 

\subsubsection*{Equations in momentum space}

For convenience, we give explicitly the equations of motion in momentum space $\vec k$. Notice $\vec k$ is the wavenumber associated to the Fourier transform of the spatial coordinates $\vec x$, not to be confused with the wavenumber associated to the Fourier transform with respect to the Lagrangian coordinates $\vec q$. For $\vec k \neq 0$, we have
\bea\label{eq:delta_LPT_fourier}
\delta(\vec k,\eta) &=& \int d^3q~e^{-i \vec k \cdot \vec z(\vec q,\eta)}\ ,\\
\Phi(\vec k,\eta) &=& -\frac{3}{2} \cH^2 \Omega_m~\frac{1}{k^2} \int d^3q~e^{-i \vec k \cdot \vec z(\vec q,\eta)}\ ,\\
\ddot  z^i(\vec q_1,\eta) + \cH \dot z^i(\vec q_1,\eta) &=&  \frac{3}{2} \cH^2 \Omega_m \int d^3 q_2 \int_k~\frac{ik^i}{k^2}  e^{i \vec k \cdot (\vec z(\vec q_1,\eta)-\vec z(\vec q_2,\eta))}\ ,\label{eom1}
\eea
where $\int_k \equiv \int \frac{d^3k}{(2\pi)^3}$.

\subsection{Effective Field Theory in Lagrangian space} 

When solving the equations of LPT perturbatively, one expands those equations assuming ${\partial s^i}/{\partial q^j}$ is small, see for example the determinant in equation \eqref{mapqx}. This ultimately sets the range of convergence of perturbation theory: the Taylor expansion around the origin cannot converge to the exact solution at a distance larger than the radius whose circle intersects the location of a singularity in the exact solution. These singularities are generically present once we include scales that have gone non-linear.  For this reason, it is not a question of being able to sum more diagrams, more terms in the series: the approach cannot work on small scales and perturbation theory can only be used to describe the long wavelength dynamics.\footnote{We present here a simple example to show why perturbation theory cannot converge to the true answer beyond the non-linear scale, no matter how many diagrams are resummed. Imagine that the solution to the equations of motion~is 
\be\label{eq:sol1}
\vec z(q,t)=\vec q-(\vec q-\vec z_c)\tfrac{t}{t_c}\ ,
\ee 
so that $\vec s(\vec q,t)=-(\vec q-\vec z_c)\tfrac{t}{t_c}$. This solution describes an in-falling  continuous set of particles that starts homogeneous at $t=0$, at time $t=t_c$ is collapsed to one point $\vec z_c$, and then at subsequent times distance themselves at constant rate. The exact solution for $\delta(\vec x,t)$ reads 
\be
1+\delta(\vec x,t)=\tfrac{1}{1-t/t_c}\ ,
\ee and clearly has a pole at $t=t_c$. Solving for the equations perturbatively in $\d s^i/\d q^j$ amounts to solving perturbatively in~$t/t_c$: $1+\delta{}^{(n)}(\vec x,t)=\sum_{i=1}^n c_i (t/t_c)^i$. Since the exact solution has a pole at $t=t_c$, the series, even as $n\to \infty$, will not converge to the true answer for $t>t_c$. 

We can also find an even more striking consequence. The time reversal solution of (\ref{eq:sol1}),
\be\label{eq:sol1n}
\vec z(q,t)=\vec q+(\vec q-\vec z_c)\tfrac{t}{t_c}\ ,\quad 1+\delta(\vec x,t)=\tfrac{1}{1+t/t_c}\ ,
\ee 
describes innocuous-looking out-flowing matter, and the exact solution for $\delta(\vec x,t)$ has no poles.  Can in this case the perturbative series converge to the exact solution? The perturbative series is now exactly the same as the one for the in-falling solution, with the replacement $t\to-t$. If we think of $t$ as a complex parameter, the failure of convergences at $t\to t_c$ for the in-falling case, due to a pole in the exact solution, means that the series will not converge beyond a circle in the complex-$t$ plane of radius $|t|=t_c$. For negative $t$, where the series describes an out-flowing solution, this implies that the perturbative series will stop converging for~$t<t_c$, even though densities are clearly becoming smaller and smaller and the exact solution is analytic! Quite counterintuitively, since the density cannot become negative, dilution of density is seen in the perturbative series as a very nonlinear event.

All of this can be verified using the exact solution of the spherical collapse. In this case, the collapsing region is equivalent to a closed FRW universe with a scale factor $a_c$. Perturbation theory amounts to Taylor expanding $a_c$ in powers of the scale factor $a$ of the external flat FRW universe. It can be checked that the series converges for $a\leq a_{\rm collapse}$, which is the value of $a$ at which the closed universe is collapsed: $a_c=0$. This means that the perturbative series will be able to converge for all values of $\delta(\vec x,a)$ as they grow all the way to infinity. At $a=a_{\rm collapse}$ the exact solution has a singularity.  However, the same perturbative series, with the replacement $a\to-a$, describes an under dense spherical region getting emptier and emptier. The perturbation series in this case will stop converge at $a=a_{\rm collapse}$. In this case, contrary to the collapsing case, $\delta(\vec x,a)$ is not yet equal to $-1$, its asymptotic value. Indeed, it can be checked that this is the case, and the perturbative series does not converge for all times beyond which $\delta(\vec x,a)\lesssim -0.7$~\cite{Sahni:1995rr}. 

Given that the true universe is surely more complicated than this, this examples clearly shows that no-resummation technique can allow us to describe the non-linear scale: perturbation theory only applies to the weakly non-linear regime.}

The idea behind LEFT is simple: we ought to construct an effective theory for the long-wavelength universe, where the short-distance physics is {\it integrated out}. As such, at a scale $k_L \ll k_{\rm NL}$, we can describe all the particles below the non-linear scale as a single point-like object endowed with new parameters that describe their extendedness (other than the mass), and whose center of mass' motion is described by the (long-wavelength) coordinate $\vec z_L(\vec q,\eta)$, with $\vec q$ representing different large regions in  Lagrangian space (see Fig.~\ref{figL1}). Clearly the dynamics of $\vec z_L(\vec q,\eta)$ is not described by \eeqref{boxed0}, but rather we need to enlarge the possible terms in the RHS due to finite size effects.  In Appendix~\ref{uvmatch} we will show that the equations in LEFT emerge from smoothing the equations in the previous section on a  scale of order the non-linear scale.\footnote{Not surprisingly, in LEFT the independence of the final result on the smoothing scale translates into the scale dependence of the multipole moments associated with each cell. See section \ref{sec:reno} for a discussion on observables in LEFT.} In section \ref{sec:action} we will derive the same equations from an action approach, following related ideas developed in \cite{nrgr1,nrgr2,nrgr3,disip1,disip2} in the context of gravitational wave emission from binaries. Here we just state the equations in LEFT without derivation, although their structure should be fairly intuitive. 

\begin{figure}[t!]
\centering
\includegraphics[width=0.6\textwidth]{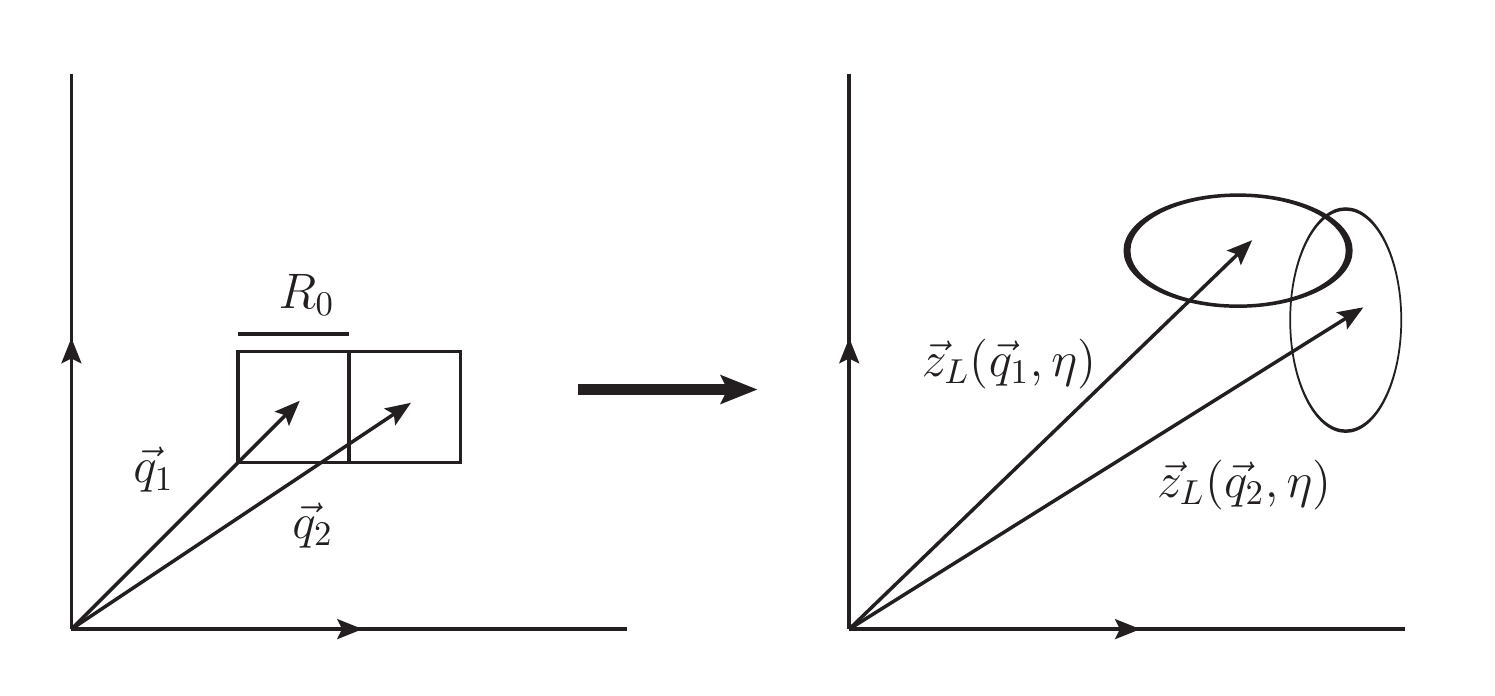}
\caption{\small Left panel: finite sized regions of size $R_0\sim 1/k_{\rm NL}$ in Lagrangian space.  Right panel: Eulerian space. The vector $\vec z_L(\vec q,\eta)$ gives the center of mass position of each Lagrangian region. Notice that upon evolution the regions will eventually overlap. See sec. \ref{sec:reno} and appendix \ref{uvmatch} for more details.}
\label{figL1}
\end{figure}

The  equation for $\vec z_L(\vec q,\eta)$ is given by:
\bea
\label{eqmotion1}
\frac{d^2 {\vec z}_L ({\vec q},\eta)}{d\eta^2} + {\cal H} \frac{d {\vec z}_L ({\vec q},\eta)}{d\eta}  &=&  - \vec\partial_x \left[\Phi_L[{\vec z}_L({\vec q},\eta)] + \frac{1}{2}  Q^{ij}({\vec q},\eta)\partial_i\partial_j \Phi_L[{\vec z}_L({\vec q},\eta)]  + \cdots \right] + \vec a_S({\vec q},\eta)\ , \nonumber \\
\eea
where $\Phi_L$ is the long-wavelength part of the potential, and the ellipses account for higher order multipole moments. All the short distance dynamics is encoded in the multipole moments that parametrize the shape of the region whose center of mass is  ${\vec z}_L(\vec q,\eta)$ and by $\vec a_S({\vec q},\eta)$, which represents an additional source of acceleration that we are now going to describe. 
The form of equation  \eeqref{eqmotion1} should be very intuitive. It describes the fact that finite sized particles with non-zero multipole moments move differently in a gravitational potential than point particles do. For example, they are sensitive not only to the gradient of the gravitational field, but also to the tidal tensor. Since we treat finite regions as point-like, by construction, the long potential $\Phi_L$ will be computed using a multipole expansion which is not valid when Lagrangian regions overlap. The acceleration $\vec a_S({\vec q},\eta)$ encodes therefore the part of the force that cannot be computed using the locations of the centers of mass ${\vec z}_L(\vec q,\eta)$ and their multipole moments, because it depends on the details of the distribution of mass inside the regions which becomes relevant when the regions overlap.  

Because we are considering large regions in Lagrangian space which develop a non-trivial shape as a result of the structure formation process, we also need to modify the Poisson equation. The resulting equation reads  (see appendix \ref{uvmatch}  and sec. \ref{sec:action} for more details): 
\beq
\label{equation2}
\partial^2_x \Phi_L = \frac{3}{2} {\cal H}^2 \Omega_m\left(\delta_{n,L}({\vec x},\eta)  + \frac{1}{2} \partial_i\partial_j  {\cal Q}^{ij}({\vec x},\eta) 
- \frac{1}{6} \partial_i\partial_j \partial_k {\cal Q}^{ijk}({\vec x},\eta) + \cdots\right)
\equiv  \frac{3}{2} {\cal H}^2 \Omega_m \delta_{m,L}({\vec x},\eta)\ .
\eeq
In the above expressions we defined the real-space {\it matter} density $\delta_{m,L}$ of the long-wavelength universe, as well as the real-space {\it number} density $\delta_{n,L}$ and real-space multipole moments:
\bea
\label{smooth1}
1+\delta_{n,L}({\vec x},\eta)&\equiv& \int  d^3{\vec q}~\delta^3( {\vec x} - {\vec z}_L({\vec q},\eta))\ , \nn \\
\label{calq}  {\cal Q}^{i_1 \dots i_p}({\vec x},\eta) &\equiv& \int d^3 {\vec q}~ Q^{i_1\ldots i_p}({\vec q},\eta)\delta^3({\vec x}-{\vec z}_L({\vec q},\eta))\ .
\eea 
To simplify the subsequent treatment, we also performed the split into irreducible representations of the rotation group:
\beq
\label{quad-lag}
Q^{ij} ({\vec q},\eta)= Q^{ij}_{\rm TF} ({\vec q},\eta)+ \tfrac{1}{3} \delta^{ij} C({\vec q},\eta),~~C({\vec q},\eta)\equiv Q^i_i({\vec q},\eta)\ , 
\eeq
where ${\rm TF}$ stands for trace free, and we introduced:
\bea
\label{smooth2}
{\cal C}({\vec x},\eta)&=&\int d^3 {\vec q}~ C({\vec q},\eta)\delta^3({\vec x}-{\vec z}_L({\vec q},\eta)) \ .
\eea
The above equations should be equally intuitive.  The number density of particles in Eulerian space is obtained by summing over all the particles at location $\vec x$.  The multipole moments in Eulerian space are obtained by summing over the multipole moments of each of these  particles.  Particles source gravity through their mass and multipoles in a standard fashion.  The matter overdensity $\delta_{m,L}$ has been defined as the source of gravity, and includes both the contribution from the change in the number density and the shapes  described by the multipole moments.\footnote{In the equations in LEFT there is no appearance of an arbitrary cutoff, as it should be since there is no cutoff dependence in physical quantities. Later on, in sections~\ref{sec:reno} and \ref{sec:oneloop}, we will perform an explicit calculation and explain how to extract physical quantities using LEFT. We can anticipate an important executive summary: when performing the loop integrals, we will need to introduce a cutoff $\Lambda$ that regularizes the integrals and precludes contributions from short distances that are not reliable within perturbation theory. This is what in Quantum Field Theory textbooks goes under the name of `regularization' (see for example Chapter 12 of~\cite{Weinberg:1995mt}). At this point, all the parameters in LEFT will acquire a cutoff dependence, e.g. $Q^{ij} ({\vec q},\eta)\to Q^{ij} ({\vec q},\eta,\Lambda)$, designed in such a way  the result of the full calculations, that is after summing the loops and the contribution coming from the multipoles, will be cutoff independent. In this case the parameters are called `bare' and the procedure is called `renormalization'. It is an essential feature, and ultimately a consistency check, that the computations in LEFT are independent of the cutoff that regularizes the loop integrals.}


\subsubsection*{LEFT in momentum space}

For convenience, we give explicitly the equations of motion in momentum space $\vec k$:
\bea\label{eq:delta_ELPT_fourier}
\delta_{m,L}(\vec k,\eta) &=& \int d^3q~{\rm exp}\left[ -i \vec k \cdot \vec z_L(\vec q,\eta) - \frac{1}{2}k^i k^j Q^c_{ij}(\vec q) + \ldots \right] \ , \label{dmkspa}\\
\Phi_L(\vec k,\eta) &=& -\frac{3}{2} \cH^2 \Omega_m~\frac{1}{k^2} \int d^3q~{\rm exp}\left[ -i \vec k \cdot \vec z_L(\vec q,\eta) - \frac{1}{2}k^i k^j Q^c_{ij}(\vec q) + \ldots \right] \ , \\  \label{eq:displacement_fourier}
\ddot  z_L^i(\vec q_1,\eta) + \cH \dot z_L^i(\vec q_1,\eta) &=& a^i_S(\vec z_L(\vec q_1,\eta)) + \frac{3}{2} \cH^2 \Omega_m \int d^3 q_2  \int_k ~\frac{ik^i}{k^2}\exp\left[i \vec k \cdot (\vec z_L(\vec q_1,\eta)-\vec z_L(\vec  q_2,\eta))\right. \nn \\ &&\nn  \left. \ \ \qquad\qquad \qquad \qquad \qquad\qquad \qquad \qquad-\frac{1}{2}k^i k^j \big(Q^c_{ij}(\vec q_1) + Q^c_{ij}(\vec q_2)\big)+ \ldots \right]. 
\eea 
Notice that by going to Fourier space, we were able to explicitly solve for the gravitational potential in terms of $\delta_{m,L}(\vec k,\eta)$. Additionally, in some of the expressions we have kept in the exponential the displacement fields. The expectation value of the exponential of a quantity can be performed as the exponential of the connected cumulants of the same quantity. For this reason, we have redefined the multipoles and kept them in the exponential.  $Q^c_{i_1\ldots i_n}$ stands for connected multipoles~\footnote{The connected part of a multipole is defined in such a way that we can write: 
\be
Q_{i_1,\ldots, i_n}=\sum_{\rm part} Q^c_{\alpha} Q^c_{\beta}\ldots\ ,
\ee
where `part' represents all possible ways to group the $n$ indices $i_1\ldots i_n$ in $\alpha,\;\beta,\ldots$ subsets, with the subsets being equal if they differ by a permutation of the indices. This definition is essentially recursive
\be
Q_{i_1,\ldots, i_n}=Q_{i_1,\ldots, i_n}^c+\sum_{\rm sub-part} Q^c_{\alpha} Q^c_{\beta}\ldots \ ,
\ee 
where `sub-part' represents all possible way to group the $n$ indices $i_1\ldots i_n$ in such a way that each set has at most $n-1$ indices.
This definition is commonly used in $S$-matrix computations in Condensed Matter and High Energy Field Theory, see the textbook~\cite{Weinberg:1995mt}, eq.~(4.3.2), for details.

}. This exponentiation may be useful if one wants to perform resummations. We discuss this briefly in section~\ref{resum}.


\subsection{Expectation values, Response \& Noise}\label{sec:response}

As in the Eulerian EFT, the short scale dynamics that determines the multipole moments and $\vec a_S({\vec x},\eta)$ can be split into different pieces: a contribution present even in the absence of long wavelength fluctuations and a response to the long wavelength perturbations.  The first piece can furthermore be split into the expectation value over short modes and a term we call intrinsic noise that accounts for the fluctuations from realization to realization. For example for $Q^{ij}$ we can write: 
\beq
\label{decomp}
Q^{ij}= \langle Q^{ij} \rangle_S+ Q^{ij}_{{\cal S}} +Q^{ij}_{{\cal R}}\ . 
\eeq
 In  $\langle Q^{ij}\rangle_S$ we take the expectation value over the short modes. The second term accounts for the fact that in each realization there is a random departure from the expectation value and the last term is the response to the long wavelength fluctuations. S stands for short, ${\cal S}$ for stochastic and  ${\cal R}$ for response. 
 
In this paper we work up to one-loop effects, so we just need to model the quadrupole moment and $\vec a_S({\vec x},\eta)$ at first order in the perturbations and in derivatives of the long-wavelength fields. For the quadrupole response for example we can write:  
\be
\label{responseI}
{\cal Q}^{ij}_{{\cal R}}(\vec z_L({\vec q},\eta),\eta)=\int d\eta^\prime~\left[A_1^{ij,lk}(\eta;\eta^\prime)\; \partial_l\partial_k \Phi_L(\vec z_L({\vec q},\eta^\prime))+A_2^{ij,lk}(\eta;\eta^\prime)\; \partial_l s_{L,k}({\vec q},\eta^\prime)+\ldots \right]\ ,
\ee
where ${\cal Q}^{ij}$ is defined in \eqref{smooth1}, and $A_i^{ij,lk}(\eta;\eta^\prime)$  are retarded Green's function that depend on the short distance dynamics. The long modes can only affect the dynamics through terms that are allowed by rotational invariance and the equivalence principle, such as $\d_i\d_j\Phi_L,\; \d_js^i_L$, etc.  

Note that in the expression in \eeqref{responseI} we did not assume locality in time as the dynamics of the short modes has a  typical times scales ${\cal O}(H^{-1})$ which is the same as that of the long modes. For this reason, the response of the quadrupole to the long wavelength perturbations depends on the field values at earlier times up to a scale of order $H^{-1}$. In this paper we will be interested in the lowest order corrections in  LEFT. We can then take the long-wavelength modes in \eeqref{responseI} at linear level. Since at linear level perturbations evolve in a $k$-independent way, the non-locality in time can be absorbed into making the response local-in-time, but with time-dependent coefficients.\footnote{ 
Using the notation from section \ref{sec:oneloop}, we can show the following. Since
\be
s^{(1)}(\vec k,\eta)\sim D(\eta)\delta_0(\vec k)\ ,
\ee
we can write
\be
Q^{ij}_{{\cal R}}({\vec q},\eta)\sim\left[\int d\eta^\prime~A_1^{ij,lk}(\eta;\eta^\prime)\; \frac{D(\eta')}{D(\eta)}\right]\partial_l\vec s_{L,k}(\vec q,\eta)\sim l^2_{ij}(\eta) \partial^i s_L^j(\vec q,\eta)\ ,
\ee
where in the last step we have implicitly defined the time-dependent parameter $l_{ij}^2(\eta)$, which can be next decomposed into trace and trace-free parts, as in \eqref{eq:quadr_response}. The situation is more subtle at higher orders. See~\cite{Carrasco:2013mua} for a treatment up to two-loops in the context of the Eulerian EFT.}  Furthermore, at linear order we can replace second derivatives of $\Phi_L$ with first derivatives of the displacement field. Clearly a displacement that is constant in space cannot lead to any response. We can therefore write, at the order at which we are working,
\bea \nn \label{eq:quadr_response}
&&\!\!\!\!\!\!\!\!Q_{ij}({\vec q},\eta) = l_S^2(\eta) \frac{1}{3} \delta_{ij} - \frac{1}{3} l^2_T(\eta) \delta_{ij}~\partial_k  s_k(\vec q,\eta) - l_{TF}^2(\eta) \left(\frac{1}{2}(\partial_i  s_j({\vec q},\eta) + \partial_j  s_i({\vec q},\eta)) - \frac{1}{3} \delta_{ij} \partial_k  s_k({\vec q},\eta)\right) \\  
&&\!\!\!\!\!\!\!\!\quad\qquad\qquad +Q^{ij}_{{\cal S}}(\vec q,\eta)\ ,
\eea
with
\beq
\langle Q^{ij}\rangle_S \equiv l_S^2(\eta) \frac{1}{3} \delta_{ij}\ .
\eeq
Similarly, the dependence on the long wavelength fields of the acceleration induced by the potential on short scales will follow the same rules. We can therefore write
\beq
\label{ashortresp}
\vec a_{S}(\vec z_L({\vec q},\eta)) = \frac{3}{2} \cH^2 \Omega_m ~l_{\Phi_S}^2(\eta) \vec \partial_q (\vec \partial_q \cdot\vec s_L(\vec q,\eta)) +\vec a_{\cal S}(\vec q,\eta)\ .
\eeq

\subsection{Renormalization, Matching \& Power Counting}\label{sec:reno}

In LEFT, the true relevant parameters of perturbation theory are manifest. It is a derivative expansion of the long modes with coefficients that are proportional to square of the short mode displacements. In fact, by inspecting \eeqref{eqmotion1} and \eeqref{equation2}, we see that the series in the higher-derivative terms can no longer be truncated when the acceleration induced by the tidal forces of the long modes over a region of the order of the random displacements becomes comparable with the acceleration of the center of mass (i.e. $\Phi_L[{\vec z}_L({\vec q},\eta)] \sim   Q^{ij}(\vec q,\eta)\partial_i\partial_j \Phi_L[{\vec z}_L({\vec q},\eta)] $), or the quadrupole moment of the mass distribution changes significantly the gravitational potential (i.e. $\delta_{n,L}({\vec x},\eta)  \sim\partial_i\partial_j  {\cal Q}^{ij}({\vec x},\eta)$). The EFT we wrote down is in Lagrangian space, and therefore $\epsilon_{s <}$ (see Fig. \ref{epsilons}) is never assumed to be small.  The additional parameters in the EFT measure the quadrupole moment of the Lagrangian mass distribution, as well as the response on short scales of the gravitational potential. 
The effect of the short distance physics is encapsulated by these new parameters. 
They indeed correct the errors induced by using an invalid perturbative approach when $k \gtrsim k_{\rm NL}$, as it happen when we perform loop integrals, so that the final result reproduces the physical contribution of the short distance physics in a correct manner. These errors from extrapolating the perturbative treatment for $k\gtrsim k_{\rm NL}$ can be finite or infinite (i.e. proportional to positive powers of a cutoff), depending on the initial power spectrum of the fluctuations. It is therefore useful to split the parameters in the sum of a potentially-divergent counter-term, and a renormalized parameter:
\beq
l^2(\eta,\Lambda) \to l^2_{\rm c.t.}(\Lambda,\eta) + l^2_{\rm ren}(\eta),
\eeq
where $\Lambda$ is the cutoff in the EFT. The theory then remains finite as $\Lambda \to \infty$ after the counter-term is properly chosen. Notice that everything we need in order to be able to describe the correct evolution of long wavelength modes is a set of (time dependent) parameters. This is a manifestation of the fact that short distance physics decouples. These parameters can be obtained directly from the UV complete theory, for our case N-body simulations, or by matching to observations.

There is always an ambiguity in the splitting of a parameter between a counter-term and renormalized part, for what the $\Lambda-$independent piece is concerned. This ambiguity is usually fixed by defining some particular renormalization conditions for the observables, for example by demanding that the computed power spectrum agrees with the observed one at a given reference wavenumber. This is usually called {\it matching} procedure. For the scope of this paper, we will mainly concern with $ l^2_{\rm c.t.}(\Lambda,\eta) $, to show that all divergencies in LPT are absorbed by the counter-terms. Of course, to obtain a physical answer, one needs to add the finite contributions as well. 

\begin{figure}[h!]
\centering
\includegraphics[width=0.6\textwidth]{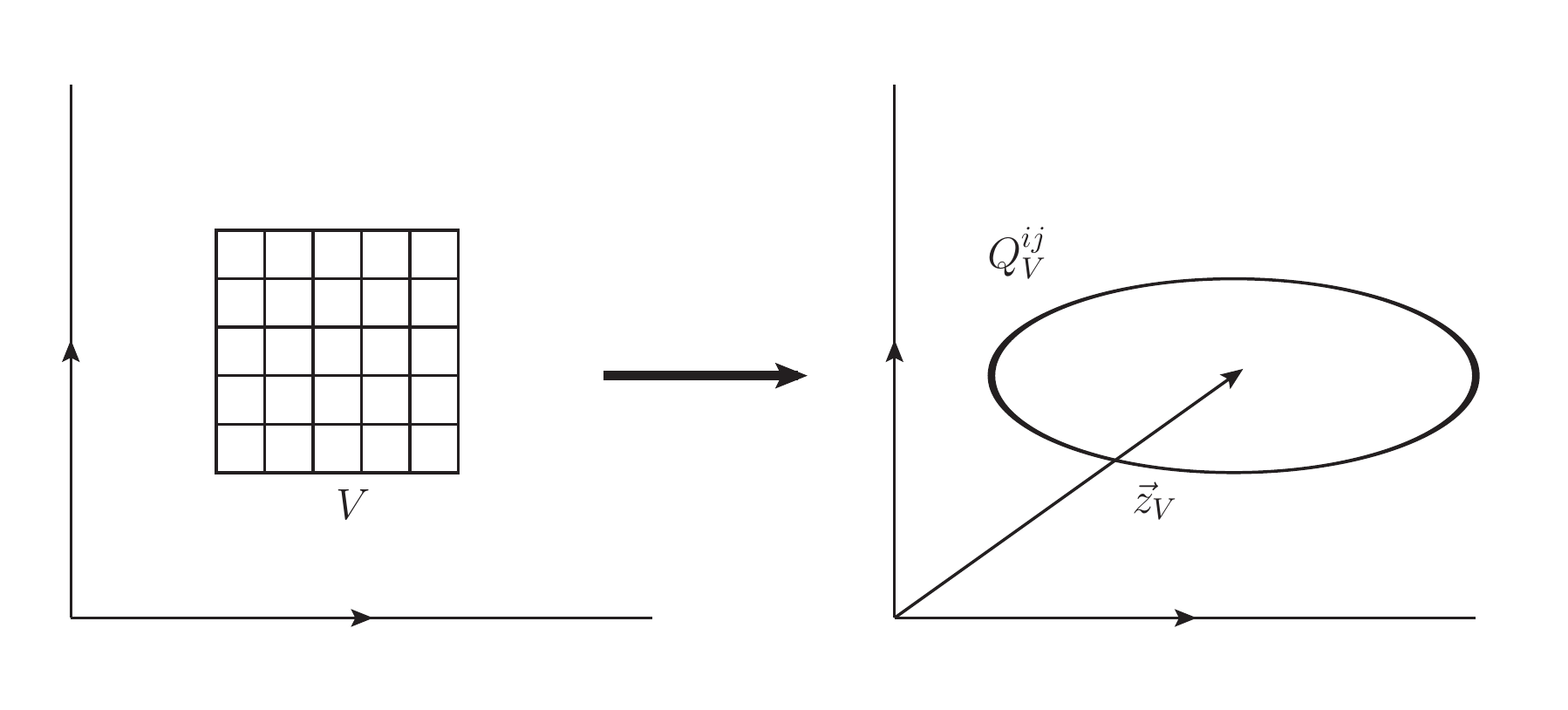}
\caption{\small Left panel: a region of size $V$ in Lagrangian space containing several cells of size $R_0 \simeq k_{\rm NL}^{-1}$, each evolving with its own quadrupole moment $Q^{ij}(\vec q,\eta)$, as shown in Fig. \ref{figL1}. Right panel: The same region in Eulerian space. The physical value of the  quadrupole moment of the entire region, $Q^{ij}_V$, must be independent of the sizes of the Lagrangian cells, i.e. it must be cutoff independent. Determining the value of the quadrupole and  other multipoles in LEFT involves free parameters that may be obtained from data or comparison with N-body simulations.}
\label{figL2}
\end{figure}

To provide more insight let us study the following example, where in the long wavelength theory we consider a certain (large) volume $V$ in Lagrangian space, whose center of mass is given by (see Fig. \ref{figL2}))
\beq
\vec z_V = \frac{1}{V} \int_V d^3 q~\vec z_L(\vec q,\eta)\ .
\eeq
The intrinsic quadrupole moment of the region is
\beq
Q_{V}^{ij} = \frac{1}{V} \int_V d^3 q~(z_L^i(\vec q,\eta)-z^i_V)(z_L^j(\vec q,\eta)-z^j_V) + Q^{ij}(\vec q,\eta) \ , 
\eeq
where we included the intrinsic quadrupole of each region described by $\vec z_L(\vec q,\eta)$. This is expected to be a well defined, measurable, object.

By writing $\vec z_L = \vec q + \vec s_L$, we have
\beq
\label{qvij}
Q_V^{ij} + z_V^i z_V^j  =\frac{1}{V} \int_V d^3 q ~\left[\left(s^i_L (\vec q,\eta) s^j_L(\vec q,\eta) + Q^{ij}(\vec q,\eta)\right) +  q^i s_L^j(\vec q,\eta)+ q^j s_L^i(\vec q,\eta) + q^i q^j\right]\  .
\eeq
By taking the expectation value on the background of the short modes (without a long-wavelength perturbation), we have
\beq
\langle Q_V^{ij} + z_V^i z_V^j \rangle_S = \frac{1}{V} \int_V d^3 q \left\langle s^i_L (\vec q,\eta) s^j_L(\vec q,\eta) + Q^{ij}(\vec q,\eta)\right\rangle_S+l_V^2{}_{ij} \ .
\eeq
where $l_V^2{}_{ij}=\int_V d^3 q\; q^i q^j$ is a geometric factor associated to the Lagrangian volume $V$.
This equation tells us two different points. First, the right-hand-side is finite (namely independent of the cutoff) and thus the left-hand-side must also be. Notice that the left-hand-side is expected to be well defined, as it represents the overall quadrupole of a given Lagrangian region. Secondly, its actual value can be obtained by measuring the analogous quantity in N-body simulations or observations (see~\cite{Carrasco:2012cv} for an implementation of this procedure in the context of the Eulerian EFT). 

Furthermore, one can consider the following correlation function of the quadrupole of a region with a far away displacement $z_L^m(\vec q_2,\eta)$:
\bea\label{eq:quad_res}
&&\left\langle z_L^m(\vec q_2,\eta ) \left( Q_V^{ij} + z_V^i z_V^j\right)\right\rangle =\\ \nonumber
&&\qquad \left\langle s_L^m(\vec q_2,\eta) \;\times\;\frac{1}{V} \int_V d^3 q \left\langle s^i_L (\vec q,\eta) s^j_L(\vec q,\eta) + Q^{ij}(\vec q,\eta)\right\rangle_{S,\vec s_L}\right\rangle\\ \nonumber
&& \quad+\left\langle s_L^m(\vec q_2,\eta) \;\times\;\frac{1}{V} \int_V d^3 q \left[q^i \left\langle s^j_L (\vec q,\eta)\right\rangle_{S,\vec s_L}+q^j \left\langle s^i_L (\vec q,\eta)\right\rangle_{S,\vec s_L}\right]\right\rangle\ .
\eea
The most interior expectation values on the right-hand-side is taken for short modes in the presence of a fixed background long mode. This expression will clearly contain terms proportional to the response of each quadrupole $Q^{ij}(\vec q,\eta)$ to the long wavelength fluctuations. Its unknown coefficient will be fixed by requiring, not only that  \eeqref{eq:quad_res} is finite, but that it also matches the analogous quantities in N-body simulations or observations. If we assume, as we will show later, that the correlation function of displacements at distance points $\langle s^i_L (\vec q_1,\eta)\; s^j_L (\vec q_2,\eta)\rangle$ has been renormalized and therefore made finite, we see that in order for the full expression to be independent of the cutoff, we must require that: 
\be
\left\langle s_L^m(\vec q_2,\eta) \left(s^i_L (\vec q,\eta) s^j_L(\vec q,\eta) + Q^{ij}(\vec q,\eta)\right)\right\rangle\ ,
\ee
is also cutoff independent. Notice that now the expectation value is on both short and long fluctuations. As we will see in the next section, implementing this procedure allows us to identify the unknown coefficients $l_S^2,\, l_T^2,\, l^2_{TF}$ in \eeqref{eq:quadr_response}. Notice that the quadruple of a given region is not independent of the coefficients $l_S^2,\, l_T^2,\, l^2_{TF}$. This is due to the fact that the quadrupole is the square of two long-wavelength fields evaluated at the same location. It is therefore sensitive to the short distance physics and needs to be renormalized. In the jargon of quantum field theory, the product of fields at the same point is often denoted as a {\it composite} operator. The quadrupole falls into this category. 


After the theory is renormalized, the remaining step in order to make predictions is to establish how many coefficients are necessary to a given order. This goes by the name of {\it power counting}, and determines at which order each term enters. By inspection of the equations of motion, it is clear that the size of the extra terms (from the quadrupole and $\vec a_S$) relative to the leading order tree-level dynamics, is suppressed by $l_{A, {\rm ren}}^2 k^2$ (for $A=(S, T,TF,\Phi_S)$). For example,  for a scaling universe with density power spectrum 
\beq \Delta^2(k) \equiv \frac{k^3}{2\pi^2} P_L(k) = \left(\frac{k}{k_{\rm NL}}\right)^{n+3},
\eeq 
they enter at order
\beq
\Delta^2_{l_{A,{\rm ren}}}(k) \propto \gamma_{A,{\rm ren}} \left(\frac{k}{k_{\rm NL}}\right)^{n+5},
\eeq
where we introduced the dimensionless parameter: $\gamma_A \equiv k_{\rm NL}^2l_A^2$, which is expected to be an order one number. In this power counting only the renormalized parameters enter, not the counter-terms. This contribution must be compared with the loop expansion, again after renormalization. This also has a well defined scaling in $k/k_{\rm NL}$. For $N$ loops, it is given by~\cite{Pajer:2013jj,Carrasco:2013sva} 
\beq
\Delta^2_{(N)}(k) \propto \left(\frac{k}{k_{\rm NL}}\right)^{(n+3)(N+1)} \quad \to\quad \Delta^2_{(1)} \propto  \left(\frac{k}{k_{\rm NL}}\right)^{2n+6}.
\eeq
It is straightforward to see that higher order effects, either from higher multipole moments, or from the same multipole moments evaluated at higher order in perturbation theory, scale as
\beq\label{eq:counter_contrib}
\Delta^2_{Q^{i_1\ldots i_n}, \vec a_S} (k) \propto  \left(\frac{k}{k_{\rm NL}}\right)^{(n+3)N+2p}, 
\eeq
with $N$ representing the number of loops, and $p$ being related to the number of extra indices of the multipoles and of derivatives in the response of $\vec a_S$. Due to locality in real space, each additional derivative will contribute with a factor $(k/k_{\rm NL})^{2p}$.

The power counting is complete with the scaling for the stochastic piece, which one can show enters at order \cite{Baumann:2010tm,Carrasco:2012cv,Pajer:2013jj,Carrasco:2013sva} 
\beq
\label{noise}
\Delta^2_{\cal S} \propto \left(\frac{k}{k_{\rm NL}}\right)^7 \ .
\eeq
and at order $(k/k_{\rm NL})^5$ for the velocity power spectrum~\cite{Carrasco:2013mua}.

Let us stress two points from the former expressions. Firstly, depending on the different values of $n$, different terms are more important than others. For example,  for $n< -1$ the correction from $\Delta^2_{l_A}(k)$ is more important than the one-loop term, provided $\gamma_A \simeq {\calo (1)}$, and both coincide for $n=-1$. Secondly, what enters in the power counting are the renormalized values for the parameters, not the (bare) potentially divergent ones. This is the case because the counter-term part is cancelled by an identical divergence in the loops. This shows that divergencies in loops do not have any particular significance: what matter is the sum of loops and counter-terms, which contributes a smaller and smaller amount as we go to higher orders. The fact that loop diagrams may be (power-law) divergent when evaluated with a cutoff in momentum space should not mislead us. Indeed, the same diagrams can be evaluated in dimensional regularization, where only logarithmic divergencies appear and the power counting is therefore simpler. We discuss dimensional regularization in appendix~\ref{app:dimreg}.

\section{Illustrative one-loop calculation}\label{sec:oneloop}

As an illustrative example in this section we show results for Einstein-de Sitter cosmologies with power law initial conditions: $P_L = A k^n$. This will allow us to see explicitly the structure of the divergences and the role of the different counter-terms in absorbing them. As stressed earlier, to obtain a final result we should add the renormalized (finite) contributions. In order to do that, the parameters of  LEFT must be fit to observations or to N-body simulations. Although this is conceptually straightforward, it goes beyond the scope of this paper (see~\cite{Carrasco:2012cv,Carrasco:2013mua} where this was carried out in the context of the Eulerian EFT). Here we show that divergencies can be absorbed by counter-terms, which provides a consistency check for LEFT. For ease of notation we drop the label $\{\}_L$ on the variables, although we will always deal with long-wavelength fields defined in the EFT. Furthermore, in what follows we will work with a cutoff regulator. We discuss dimensional regularization in appendix \ref{app:dimreg}, and in particular the case $n=-1$.

\subsection{Basics}
In this section we follow the notation in \cite{Matsubara:2008wx}. We can solve iteratively in $\d_i s_j$ the equations for the displacement. Working in $q$-variables Fourier space, we find
\beq
\label{vecsn}
\vec s^{(n)}(\vec k,\eta) = \frac{iD(\eta)^n}{n!} \int_{\vec p_1}\ldots \int_{\vec p_n} (2\pi)^3\delta^3\left(\vec k_t - \vec k\right)~{\vec L}^{(n)}(\vec k_1\ldots \vec k_n) \delta_0(\vec k_1)\ldots \delta_0(\vec k_n),
\eeq
where $\delta_0 \equiv \delta(\eta=\eta_0)$, $\vec k_t = \sum_i^n \vec k_i$, $D(\eta)$ is the growth factor (normalized to $D(\eta_0)=1$), and $\vec L^{(n)}$ is real. Up to $\vec L^{(3)}$ we have~\cite{Matsubara:2008wx}
\bea
\vec L^{(1)} &=& \frac{\vec k}{k^2} \\
\vec L^{(2)} &=& \frac{3}{7} \frac{\vec k_t}{k_t^2} \left(1- \frac{(\vec k_1\cdot \vec k_2)^2}{k^2_1k^2_2}\right)\label{l2s2}\\
\label{l3s3} \vec L^{(3)} &=& \frac{5}{7} \frac{\vec k_t}{k_t^2} \left(1- \frac{(\vec k_1\cdot \vec k_2)^2}{k^2_1k^2_2}\right)
\left[ 1- \left\{\frac{(\vec k_1+\vec k_2)\cdot\vec k_3}{|\vec k_1+\vec k_2|k_3}\right\}^2\right] \nn \\ &-& \frac{1}{3}
 \left(1- 3\frac{(\vec k_1\cdot \vec k_2)^2}{k^2_1k^2_2}+2 \frac{(\vec k_1\cdot\vec k_2)(\vec k_3\cdot\vec k_2)(\vec k_1\cdot\vec k_3)}{k_1^2k_2^2k_3^2}\right) + \vec k_t \times \vec T,
\eea
with $\vec T$  a transverse piece which does not contribute at one-loop~\footnote{The reason is simple, only $\vec s^{(1)} \propto \vec k$ is available to correlate with $\vec s^{(3)}$ at one-loop order, and obviously $(\vec k\times \vec T)\cdot \vec k=0$.}. 
From here we can compute:
\bea
\langle \vec s_i(\vec k_1) \vec s_j(\vec k_2)\rangle &=& -(2\pi)^3 \delta^3(\vec k_1+\vec k_2) C_{ij}(\vec k_1,\vec k_2)\ ,\\
\langle \vec s_i(\vec k_1) \vec s_j(\vec k_2)\vec s_l(\vec k_3)\rangle &=& +i (2\pi)^3 \delta(\vec k_1 + \vec k_2 + \vec k_3)  C_{ijl}(\vec k_1,\vec k_2,\vec k_3)\ ,
\eea
which entails convoluted integrals with ($A \equiv A(\eta) = A_0 D(\eta)^2$)
\beq
P_L = A k^n = 2\pi^2 \frac{k^n}{k_{\rm NL}^{n+3}(\eta)}.
\eeq
Hence 
\bea
C_{ij}^{(11)}(k) &=&- \frac{k_i k_j}{k^4} P_L(k)\ , \\
C_{ij}^{(22)}(k) &=& -\frac{9}{98} \frac{k_i k_j}{k^4} Q_1(k)\ , \\
C_{ij}^{(13)}(k) &=& C_{ij}^{(31)}(k)=- \frac{5}{21} \frac{k_i k_j}{k^4} R_1(k)\ , \\ \nonumber
\int_p C_{ijl}^{(112)}(k,-p,p-k) &=& \int_p C_{ijl}^{(121)}(k,-p,p-k) = \frac{3}{14} \left( -\frac{k_i k_j k_l}{k^6} (R_1(k) + 2 R_2(k)) + \delta_{jl}\frac{k_i}{k^4} R_1(k)\right), \\ 
&&\\
\int_p C_{ijl}^{(211)}(k,-p,p-k) &=& \frac{3}{14} \left(- \frac{k_i k_j k_l}{k^6} (Q_1(k) + 2 Q_2(k)) + \delta_{jl}\frac{k_i}{k^4} Q_1(k)\right)\ ,
\eea
where the functions $R_{1(2)}(k), Q_{1(2)}(k)$ are defined in \cite{Matsubara:2008wx}. The terms that depend upon $Q_{1(2)}$ introduce divergences that scale
like $k^7 \Lambda^{2n-1}$ and therefore are handled by the stochastic terms (see \eeqref{noise}) \cite{Baumann:2010tm,Carrasco:2012cv,Pajer:2013jj,Carrasco:2013sva}. (This divergence is not present for example for $n=-1$.) The remaining divergences,  depending on $R_{1(2)}$,
\beq
R_{1(2)} (k) = P_L(k) \frac{k^3}{4\pi^2} \int_0^\infty dr P_L(kr) \tilde R_{1(2)}(r)\ ,
\eeq
with
 \beq
 \tilde R_1(r \to \infty) \to \frac{16}{15},~~ \tilde R_2(r \to \infty) \to -\frac{4}{15} \label{limitr}\ ,
\eeq
are the ones regularized by the background expectation value and response of the quadrupole moment $Q^{ij}(\vec q,\eta)$, and $\vec a_{S,\cal R}(\vec q,\eta)$, which we discuss next.

Let us emphasize that nothing prevents us from studying the cases for which SPT gives finite results and the above integrals are finite as $\Lambda \to \infty$. However, the contribution from the UV modes must still be corrected by a counter-term despite the latter remaining finite as we remove the cutoff. An EFT is the natural path to systematically parameterize the short distance dynamics.

\subsection{Quadrupole moment \& composite operators\label{sec:quad}}

As we described in sec. \ref{sec:reno}, we require the combination
\beq
\label{compositeqij}
s_i (\vec q,\eta) s_j(\vec q,\eta) + Q_{ij}(\vec q,\eta)\ ,
\eeq
to be cutoff independent, both when we take the expectation value on the short modes, and also when we correlate it with a point far away. As we discussed before, this is a composite operator. We have divergencies both at the level of the expectation value
\beq\label{eq:backround_quad}
\langle s_i(\vec q,\eta) s_j(\vec q,\eta)\rangle \propto \int_p C_{ij}(p)\ ,
\eeq
and at the level of correlation with the displacement at a far away point. In $q$-Fourier space, by calling $\vec q = \vec q_1-\vec q_2$, we have 
\bea\label{eq:quad_response_div}
&&\langle s_l( \vec q_2,\eta) s_i(\vec q_1,\eta) s_j(\vec q_1,\eta)\rangle\quad \to \\ \nonumber
&&\qquad\left\langle s_l(\vec k,\eta) \int_p s_i(\vec p-\vec k,\eta) s_j(-\vec p,\eta)\right\rangle' = i \int_p C^{(112)}_{ijl} ( k,-p,p-k) + C^{(121)}_{ijl}( k,- p, p- k)\ ,  
\eea
where the $\langle\ldots\rangle'$ means that we have removed a factor of $(2\pi)^3\delta^{(3)}(\vec k+\vec k')$ from the expectation value. 
The divergence of the first kind in \eeqref{compositeqij} will be renormalized by the counter-term  associated to the expectation value of the quadrupole: $l_S^2 \equiv \langle Q^i_i\rangle$; while the second kind of divergence from \eeqref{eq:quad_response_div} will be renormalized by the response of the quadrupole
\beq
\label{response}
Q_{ij,{\cal R}} = -\frac{1}{3} l_T^2 \delta_{ij}~\partial_k  s_k - l_{TF}^2 \left(\frac{1}{2}(\partial_i  s_j + \partial_j  s_i) - \frac{1}{3} \delta_{ij} \partial_k  s_k\right)\ . 
\eeq
Using \eeqref{eq:quadr_response} for a scaling power spectrum of the form $P_L = Ak^n$, \eeqref{response} becomes:
\beq
\label{ctqij}
\langle Q_{ij,{\cal R}} s_l \rangle (k) = i P_L(k) k^2 \left( \frac{k_l}{k^4} \frac{1}{3} (l_{T}^2-l_{TF}^2) \delta_{ij} + l_{TF}^2 \frac{k_i k_j k_l}{k^6}\right). 
\eeq

As we mentioned, the purpose of this section is to show that the divergencies from the composite operators or from the LPT loops are absorbed by the counter-terms of LEFT. For this reason, we concentrate only on the divergent part of these expressions. Using \eeqref{limitr}, we obtain:
\bea
&& R^{\Lambda}_1 (k) = \frac{16}{15(n+1)} P_L(k) \frac{k^3}{4\pi^2} A k^n \frac{\Lambda^{n+1}}{k^{n+1}} \equiv \frac{8}{15}  k^2 P_L(k) l_\Lambda^2(\Lambda,\eta)\ , \\
&& R^{\Lambda}_2(k) = -\frac{1}{4} R^{\Lambda}_1 (k)\ ,
\eea
where we defined
\bea
&& l_\Lambda^2(\Lambda,\eta) \equiv\frac{1}{2\pi^2}\int^\Lambda_0 dp\; P_L(p) = \frac{\Lambda^{n+1}}{k_{\rm NL}^{3+n}(n+1)}\ . 
\eea
Notice that we have included a superscript $\Lambda$ to $R_{1,2}$ to stress that we are keeping only the divergent part. For illustrative purposes, and to simplify the computations, we restrict ourselves to study scaling universes, $P_L = Ak^n$, with $-1 < n<1$, for which the above are the only divergent pieces. (In appendix \ref{app:dimreg} we also study the case $n=-1$). Of course there are also finite corrections to those integrals, which we do not include in this analysis since we are only concerned about the divergent parts~\footnote{It is worth stressing once again the following. For $-3<n<-1$, the $R_{1,2}$ integrals are UV convergent. This does not mean that the new terms from LEFT should not be added. In fact, the new terms in LEFT contribute directly at a given order in $k/k_{\rm NL}$, as shown in \eeqref{eq:counter_contrib}, and must be included. They represent the the influence of the short distance physics on long scales and must be systematically included.}.

Let us therefore proceed with the renormalization of the quadrupole. For the background expectation value \eeqref{eq:backround_quad}, we impose:
\beq
\langle s_i (\vec q,\eta) s_j(\vec q,\eta)\rangle + l^2_S \frac{\delta_{ij}}{3} \to {\rm finite}.
\eeq
Since
\beq
\langle s_i (\vec q,\eta) s_j(\vec q,\eta)\rangle =  \frac{1}{3} \delta_{ij} A \int^\Lambda_0 dk \frac{4\pi^2}{(2\pi)^3} k^n= \frac{1}{3} \delta_{ij} \int^\Lambda_0 dk \frac{k^n}{k_{\rm NL}^{n+3}(\eta)}=\frac{1}{3} \delta_{ij} l_\Lambda^2(\Lambda,\eta) \ ,
\eeq
then, splitting $l_S = l_{S,\rm ren}(\eta) + l_{S,\rm c.t.}(\Lambda,\eta)$, we require:
\beq
\label{lsct}
l_{S,\rm c.t.}^2(\Lambda,\eta) = - l_\Lambda^2(\Lambda,\eta)+ {\rm finite}\ .
\eeq

For the response in \eeqref{eq:quad_response_div}, adding both contributions from $C^{(112)}_{ijl}$ and $C^{(121)}_{ijl}$, we obtain (again after Fourier transform)
\bea
&& \left\langle s_l(\vec k,\eta) \int_p s_i(\vec p-\vec k,\eta) s_j(-\vec p,\eta)\right\rangle' = i \int_p C^{(112)}_{ijl} (k,-p,p-k) + C^{(121)}_{ijl}(k,-p,p-k) \\
&=& i \frac{3}{7} \left( \frac{k_i k_j k_l}{k^6} (R_1 + 2 R_2) - \delta_{ij}\frac{k_l}{k^4} R_1\right) \quad \toUV\quad 
\quad (i  k^2 P_L(k) ) \frac{4}{35} l_\Lambda^2(\Lambda,\eta) \left( -\frac{k_i k_j k_l}{k^6} +  2\delta_{ij}\frac{k_l}{k^4}\right),  \nn
\eea
where in the last line we extracted the UV limit of the expression to isolate the divergences. By using \eeqref{ctqij}, and splitting again into renormalized part and a counter-term, we see that in order for the composite operator (\ref{compositeqij}) to have finite correlation with far away displacements, we require
\bea\label{eq:counter_sol}
l_{TF,{\rm c.t.}}^2(\Lambda,\eta) &\equiv& \frac{4}{35} l_\Lambda^2 (\Lambda,\eta)\label{ltfct}\ ,\\ \nn
l_{T,{\rm c.t.}}^2(\Lambda,\eta) &\equiv&-\frac{4}{7} l_\Lambda^2(\Lambda,\eta)\label{ltct}\ .
\eea

This result shows that the quadrupole moment of the long wavelength theory can be regularized by a proper choices of counter-terms. We discuss next the regularization of the displacement field. We will show that the values obtained previously are consistent with the values of the counter-terms required to regularize the two-point function of the displacement field. 


\subsection{Displacement\label{sec:displacement}}

Let us now perform  the renormalization of the displacement fields.  It is useful to define the {\it source}~$\vec{\cal S}(\vec q,\eta)$, which will source the perturbative corrections order by order. From \eeqref{eq:displacement_fourier}, we have
\beq
\label{source}
\ddot  s^i(\vec q_1,\eta) + \cH \dot s^i(\vec q_1,\eta) + \frac{3}{2} \cH^2 \Omega_m \int {d^3 k \over (2 \pi)^3}   {i k^i \over k^2 } \int d^3 q_2 \;e^{ i \vec k \cdot (\vec q_1 -\vec  q_2)}\;  i\,k^j \left(s_j(\vec q_2,\eta)-s_j(\vec q_1,\eta)\right) \equiv {\cal S}^i(\vec q_1,\eta)\ .
\eeq
By solving perturbatively in the fluctuations, we have that the $n$-th order solution takes the form
\beq
\vec s^{(n)}(\vec q,\eta) = \int d\eta^\prime G(\eta,\eta^\prime) \vec{\cal S}^{(n)}(\vec q,\eta^\prime)\ ,
\eeq
where $G(\eta,\eta')$ being the Green's function associated with the linear equation of motion. ${\cal S}^{(n)}$ is the $n$-th order source.  It should be made clear that the $n$-th order source is made of displacement fields, possibly evaluated at some non-linear order, as well as the new terms from LEFT. As we see in \eeqref{eq:counter_contrib}, they contribute at a given non-linear order.  From here it is clear that, in order to make the correlation of the displacement finite, all we require are a finite correlation between $\vec{\cal S}$ and $\vec s$ at distant points. We focus next directly on those correlations.

Since in this paper we are limiting ourselves to the first non-trivial order, we can expand the source up to cubic order:
\bea
&&{\cal S}^l(\vec q_1) = a^l_S(\vec q_1) +  \frac{3}{2} \cH^2 \Omega_m  \int {d^3 k \over (2 \pi)^3}   {i k^l \over k^2 } \int d^3 q_2 \exp[ i \vec k \cdot (\vec q_1 -\vec  q_2)] \times \\ 
&&\qquad\qquad\times \left\{  \tfrac{1}{2}(ik_i)(ik_j)\big[ s_i(\vec q_1) s_j(\vec q_1) +Q_{ij}(\vec q_1)+s_i(\vec q_2) s_j(\vec q_2)+ Q_{ij} (\vec q_2)- 2 s_i(\vec q_1) s_j(\vec q_2)\big]  \right.   \nn \\ 
&&\qquad\qquad\qquad +   \frac{1}{6}(i k_i)(i k_j) (i k_r) \left[ 3 s_r(\vec q_1) (s_i(\vec q_2) s_j(\vec q_2)+Q_{ij}(\vec q_2))- 3 s_r(\vec q_2) (s_i(\vec q_1) s_j(\vec q_1)+Q_ {ij}(\vec q_1))\right.\nn\\ 
&&\qquad\qquad\qquad+ \left.\left.  \left(s_i(\vec q_1) s_j(\vec q_1) +3 Q_{ij}(\vec q_1)\right) s_r(\vec q_1) - \left(s_i(\vec q_2) s_j(\vec q_2) +3 Q_{ij}(\vec q_2)\right) s_r(\vec q_2) \right]\right\} + \ldots \nn \label{forcep}
\eea
Here we suppressed the time-dependence of the variables for simplicity. We immediately recognize the combination $s_i(\vec q_2) s_j(\vec q_2)+Q_{ij} (q_2)$, which leads to finite results provided we choose the counter-terms as shown in the previous section. Indeed the factor of $3$ in the last two terms precisely accounts for the three different ways to correlate each $\vec s(\vec q_2)$ with a far away point $\vec s(\vec q_3)$.

There are, however, remaining terms that are potentially divergent which are not fixed by the background expectation value and response of the quadrupole moment. The first such term is easily identified as the last one in the second line of \eqref{forcep}, which reads, after correlating with $\vec s(\vec q_3)$ and transforming into Fourier space,
\bea
\label{newint}
&&\langle {\cal S}_l(\vec q_1) s_m(\vec q_3)\rangle \supset\nn \\ 
&&\nn \quad \langle \left({\cal S}_l(\vec q_1)\right)_{2} s_m(\vec q_3)\rangle'=-  \frac{3}{2}\cH^2 \Omega_m \int_k   {i k_l \over k^2 } (ik_i)(ik_j) \int d^3 q_2\; e^{ i \vec k \cdot (\vec q_1 -\vec  q_2)}\,  \langle s_i(\vec q_1) s_j(\vec q_2) s_m(\vec q_3) \rangle \nn \\ &&\quad \Rightarrow\quad  \langle ({\cal S}_l)_{2}s_m\rangle'= -\frac{3}{2}\cH^2 \Omega_m \int_p  {i p_l \over p^2 } (ip_i)(ip_j)\; i\, C_{ijm}(p,-k,k-p)\ ,
\eea
where the subscript $_{2}$ in ${\cal S}$ underlines that we are taking in ${\cal S}$ a term that is squared in the displacement fields.
This term is divergent and cannot be absorbed into the multipole moments $l_{TF}^2,\,l_{T}^2, l_S^2$, since, as discussed in sec.~\ref{sec:reno}, those counter-terms have been already fixed by the renormalization of the composite operator $s^is^j+Q^{ij}$, which is required to make the multipole moment of a given region in Lagrangian space finite.

Note indeed that the integral in \eeqref{newint} is slightly different from what we found before for the renormalization of $s_i s_j$, since in this case the composite operator involves derivatives (this is shown by the factors of $p$ inside the $dp$ integral). The above term is equivalent to taking a correlation of the displacement with the following new contact  operator  
\be 
{\cal S}_l\supset \calo_l (\vec q_1) \equiv  s_j(\vec q_1) { \partial_l \partial_j \partial_k\over \partial^2} s_k(\vec q_1)\ , 
\ee
as it can be easily checked by going to real space. This new contact operator needs to be regularized in order to make the displacement finite.  Let us compute the divergent contribution. To the order we work here we can isolate it as follows. We first split the displacement in a gradient and a curl term as
\be
\begin{split}
\vec s =& \  \vec s_{\parallel} + \vec s_{\perp} \ , \\
\vec s_{\parallel}= \vec \partial \psi \ , & \ \ \vec\partial\cdot \vec s_{\perp} =0\ .
\end{split}
\ee
In this way we have
\be
\begin{split}
s_j(\vec q_1) { \partial_l \partial_i \partial_j \over \partial^2} s_j(\vec q_1) &=  s_j(\vec q_1) \partial_l s_{j}(\vec q_1) -  s_j(\vec q_1) \partial_l s_{\perp,j}(\vec q_1) \\
&= \frac{1}{2}\partial_l [  s_j(\vec q_1) s_j(\vec q_1)]  - \partial_l [  s_{\perp,j}(\vec q_1) s_{\perp,j}(\vec q_1)]- s_{\parallel,j}(\vec q_1) \partial_l s_{\perp,j}(\vec q_1)\ .
\end{split}
\ee
Since the divergent terms in LPT come from $C^{(121)}_{ijl}$ and $C^{(211)}_{ijl}$ to this order (it can be shown $C^{(112)}_{ijl}$ does not lead to a divergence, as expected), the curl piece of the displacement (which first enters in $\vec s^{(3)}(\vec q,\eta)$ in \eqref{vecsn}) does not contribute. Hence at one-loop the divergence from the $( \vec{\cal S})_{2}$ term in \eqref{newint} becomes:
\bea
&&\langle \left({\cal S}_l\right)_{2}\;s_r\rangle' = - \frac{3}{2}\cH^2 \Omega_m \int_p^\Lambda   {i p_l \over p^2 } (ip_i)(ip_j)~i~C_{ijr}(p,-k,k-p)\nn\\
&&\qquad \qquad\toUV\quad -   \frac{3}{2}\cH^2 \Omega_m~(ik^l)\int^\Lambda_p ~i ~ \tfrac{1}{2} C_{iir}(p,-k,k-p) =  \frac{3}{2}\cH^2 \Omega_m  k^2 P_L(k)  \frac{2}{7} l_\Lambda^2(\Lambda) \frac{k^l k^r}{k^4}  \nn \\
&&\  \qquad\qquad\qquad=  \frac{3}{2}\cH^2 \Omega_m k^2 C_{lr}^{(11)}(k) \left( \frac{2}{7} l_\Lambda^2(\Lambda)\right).
\eea

There is another source of divergences in computing $\langle {\cal S}_l(\vec q_1) s_m(\vec q_3)\rangle$. This comes from the cubic terms in the third line of~\eqref{forcep}, when we correlate $\vec s(\vec q_3)$ with one of the $\vec s(\vec q_{1(2)})$ inside the parenthesis~\footnote{In the fourth line all the divergent terms are nicely accounted by the background expectation value of the quadrupole moment. Note there are 3 possible way to contract with a far away point.}. These `crossed' terms, for instance
\beq
\int {d^3 k \over (2 \pi)^3}   {i k^l \over k^2 } \int d^3 q_2 \exp[ i \vec k \cdot (\vec q_1 -\vec  q_2)](i k_i)(i k_j) ( ik_r)\langle s_m (\vec q_3) s_i (\vec q_2)\rangle \langle s_r (\vec q_1) s_j(\vec q_2)\rangle\ ,
\eeq
are clearly not regularized by the background expectation value of the quadrupole moment, which cancels the divergence generated when we contract 
\beq
\langle s_m (\vec q_3) s_r (\vec q_1)\rangle \langle s_i (\vec q_2) s_j(\vec q_2) + Q_{ij}(\vec q_2)\rangle \to~{\rm finite}\ .
\eeq
Similarly the response of the quadrupole will not enter here, as it would contribute to higher order. 
The remaining divergent contribution to the correlation function thus becomes
\beq
\langle ({\cal S}_l)_3\;s_m\rangle (k) =  \frac{3}{2}\cH^2 \Omega_m  C^{(11)}_{im}(k) \int_p^\Lambda \left[  {(p_l-k_l)(p_r-k_r)(p_i-k_i)(p_j-k_j) \over |\vec p-\vec k|^2 } - {p_l p_ip_j p_r \over p^2 }  \right] C_{jr}^{(11)}(p)\ ,
\eeq
where the the subscript $_3$ underlines that we are taking a term in ${\cal S}$ that is cubic in the displacements.
The divergent part can be evaluated to give~\footnote{Let us give some details of the evaluation. We first write
\be\label{eq:general_form}
\int_p^\Lambda \left[  {(p_l-k_l)(p_r-k_r)(p_i-k_i)(p_j-k_j) \over |\vec p-\vec k|^2 } - {p_l p_ip_j p_r \over p^2 }  \right] C_{jr}^{(11)}(p)=l_{\Lambda^2}\left[\alpha\, k_l k_i+\beta\, k^2\delta_{ij}\right].
\ee
By contracting \eeqref{eq:general_form} with $\delta^{il}$, the integral simplifies remarkably and we obtain $\alpha+3\beta=\tfrac{1}{3}$. A second equation is provided  by contracting \eeqref{eq:general_form} with $k^l k^i$. We obtain
\be
k^4(\alpha+\beta)=\int_p^\Lambda \frac{P_L(p)}{p^4}\left[\frac{(\vec p\cdot\vec k-k^2)(p^2-\vec k\cdot\vec p)^2}{(\vec p-\vec k)^2}-p^2 (\vec k\cdot\vec p)^2\right]\ .
\ee
Since we are interested in the divergent part, we can neglect all terms that contribute to an order in $k$ higher than $k^4$ as $k\to 0$. The resulting expression is quite simple, and it can be easily handled. This leads to the second equation $\alpha+\beta=\tfrac{13}{15}$. Solving for $\alpha$ and $\beta$ we obtain \eeqref{eq:div_3}.
}
\beq\label{eq:div_3}
\langle ({\cal S}_l)_3\;s_r\rangle (k)\quad \toUV \quad  \frac{3}{2}\cH^2 \Omega_m  k^2 C^{(11)}_{lr}(k) \left(\frac{13}{15} l_\Lambda^2(\Lambda)\right) \ .
\eeq

The above analysis suggests we need a new counter-term. Indeed, this is represented by the term $\vec a_S$, which accounts for the response of the potential on short scales to the presence of a long-wavelength perturbation. From \eeqref{ashortresp}, we have
\beq
\vec a_{S,\cal R}(\vec q,\eta) = \frac{3}{2} \cH^2 \Omega_m~l_{\Phi_S}^2(\eta) \vec \partial_q \vec \partial_q \cdot \vec s(\vec q,\eta) + \ldots\ .
\eeq
This new term will provide a counter-term, $l_{\Phi_S,\rm c.t.}$, which we need to judiciously choose to cancel the extra divergences. Hence, from
\beq
\langle a_{S,\cal R}^i\; s_j\rangle_{\rm c.t.}(\vec k) = \frac{3}{2} \cH^2 \Omega_m l_{\Phi_S,\rm c.t.}^2(\Lambda) (ik^i) (ik^l) C^{(11)}_{lj}(k) = -\frac{3}{2} \cH^2 \Omega_m  l_{\Phi_S,\rm c.t.}^2(\Lambda) k^2 C^{(11)}_{ij}(k)\ ,
\eeq
we can choose
\beq
 l_{\Phi_S,\rm c.t.}^2(\Lambda,\eta) = \left(\frac{13}{15}+\frac{2}{7}\right)~ l_\Lambda^2(\Lambda,\eta) = \frac{121}{105}l_\Lambda^2(\Lambda,\eta)\ ,
 \eeq
 where we reinstalled the time-dependence in the arguments for clarity sake. In this way, both
the correlation for the displacement and the quadrupole of large Lagrangian regions are made finite.

\subsection{Combined parameters}\label{sec:comb}

We have seen that all the parameters in LEFT, such as $l_S,\,l_{T},\,l_{TF},l_{\Phi_S},\ldots$ can be obtained  through a detailed matching procedure, that includes for instance measuring the quadrupole moment of a region in Lagrangian space and the power spectrum of the displacement fields. However, if one is solely interested in calculating some specific correlation functions, for example the correlation function of the displacement field or of the density field, then some of these parameters will enter in different linear combinations, unless the fields are connected by a conservation law. Hence, for the correlation of the density and the displacement fields, which are not connected by a conservation law, we expect that only two parameters are necessary~\footnote{The same situation occurs in the Eulerian EFT up to two-loops~\cite{Carrasco:2013mua}. If one is interested in correlation functions of the density field, only one parameter is necessary. The same parameter enters in the computation of correlations involving the momentum field, since these two fields are connected by matter conservation.  However, if one is interested also in velocity correlations, since the velocity field is a composite operator in the Eulerian EFT, it requires a different parameter \cite{Mercolli,Carrasco:2013mua}.}.

Let us obtain an equation directly for $\vec s(\vec q,\eta)$. Since at this order $\vec s_{\perp}$ can be neglected, we can actually work directly with the scalar quantity $\theta(\vec q,\eta) = \vec \partial_q\cdot \vec s(\vec q,\eta)$. We have, using \eqref{eqmotion1}, \eqref{equation2}, \eqref{eq:quadr_response} and  \eqref{ashortresp},
\bea
\ddot \theta(\vec q,\eta) + \cH \dot\theta(\vec q,\eta) &=& -\partial_x^2\Phi - \frac{l_S^2}{6}\partial_x^4\Phi + \frac{3}{2} \cH^2 \Omega_m  l_{\Phi_S}^2~  \partial^2_q  \theta(\vec q,\eta)+\ldots\\
\partial^2_x\Phi &=& -\frac{3}{2}\cH^2\Omega_m \left( \theta(\vec q,\eta) + \frac{1}{6} (l_S^2 + l_T^2 + 2l_{TF}^2) \partial_q^2 \theta(\vec q,\eta) +\ldots \right)\label{poisdel}\ ,
\eea
from which we derive
\bea
\ddot \theta(\vec q,\eta) + \cH\dot \theta(\vec q,\eta) -\frac{3}{2}\cH^2\Omega_m \theta(\vec q,\eta) &=& \frac{3}{2}\cH^2\Omega_m  \frac{1}{6}\left( l_T^2 + 2 l_{TF}^2 +2l_S^2+ 6l_{\Phi_S}^2\right) \partial_q^2 \theta(\vec q,\eta) +\ldots \nn \\ \label{thetaeq}
&\equiv& \frac{3}{2}\cH^2\Omega_m l^2_{s,\rm comb}\partial_q^2 \theta(\vec q,\eta) +\ldots\ ,
\eea
where we defined
\beq
l^2_{s,\rm comb} \equiv \frac{1}{6}\left( l_T^2 + 2 l_{TF}^2 +2l_S^2+ 6l_{\Phi_S}^2\right)\ .
\eeq 
To arrive to these expressions we used \beq \delta_n(\vec x,\eta) = - \theta(\vec q,\eta)+ \ldots,\label{delft}\eeq and the ellipses include higher-order terms in the expansion between density and displacement and the relation between $\vec x$ and $\vec q$ derivatives, as well as higher dimensional operators from the new terms in LEFT.

From the one-loop result for the correlation of the displacement we know that 
\beq\label{eq:31lpt}
\langle \theta^{(1)}\theta^{(3)}\rangle' = k^i k^j C^{(13)}_{ij}(k) = \frac{5}{21} R_1(k) \quad\toUV\quad \frac{5}{21} R^{UV}_1(k) = \frac{8}{63} l_\Lambda^2 (k^2 P_L(k))\ ,
\eeq
which we can be regularized with a proper choice of the counter-term $l^{\rm c.t.}_{s, \rm comb}$. To obtain the contribution proportional to $l^{\rm c.t.}_{s, \rm comb}$, we will  solve the equation in \eqref{thetaeq} using the Green's function for an Einstein-de Sitter universe,
\bea
&&G(a,a')=-\frac{2}{5}\frac{1}{{\cal H}^2 a} \left[\frac{a}{a'}-\left(\frac{a}{a'}\right)^{-3/2}\right]\ ,
\\ \nonumber
&&\qquad \Rightarrow\quad\int d a^\prime G(a,a^\prime) (a^\prime)^n = \frac{a^n}{n(n+5/2)\cH^2},~{\rm where}~~~\cH = 2/\eta,~\Omega_m=1,~ a/a_0 = \eta^2/\eta_0^2 \ . 
\eea
The time dependence of the counter term must match the one of~\eeqref{eq:31lpt}, which is $D(a)^4$, where $D(a)$ is the growth factor. Since the counter-term is evaluated at linear order, we need
\beq
l^{\rm c.t.}_{s, \rm comb}(\Lambda,\eta) \propto l_\Lambda(\Lambda,\eta) \propto D^2(\eta) = \frac{a^2}{a_0^2}\ ,
\eeq
where $a_0$ is the present day scale factor, that we often set to be equal to $a_0=1$.
Then, writing $l^{\rm c.t.}_{s, \rm comb}(\Lambda,\eta) =  l^{\rm c.t.}_{s, \rm comb}(\Lambda,\eta_0) a^2(\eta)/a_0^2$, we obtain
\beq
\theta_{\rm c.t.}(\vec q,a) = \frac{1}{6} l^{\rm c.t.}_{s, \rm comb}(\Lambda,\eta_0){}^2 \,\frac{a^2}{a_0^2}\,\partial^2 \theta^{(1)}(\vec q,a)\ ,
\eeq
where we used $\theta^{(1)} \propto a(\eta)$. From here,
\beq
\langle \theta^{(1)} \theta_{\rm c.t.}\rangle =- \frac{1}{6}l^{\rm c.t.}_{s, \rm comb}(\Lambda,\eta_0){}^2\; k^2 P_L(k) \;a(\eta)^4\ . 
\eeq
Therefore, we can cancel the divergence by taking 
\beq\label{eq:lscomb}
\frac{1}{6}l^{\rm c.t.}_{s, \rm comb}(\Lambda,\eta_0){}^2 =  \frac{1}{36}\left( l_T(\Lambda,\eta_0)^2 + 2 l_{TF}(\Lambda,\eta_0)^2 +2l_S(\Lambda,\eta_0)^2+ 6l_{\Phi_S}(\Lambda,\eta_0)^2\right)_{\rm c.t.} =\frac{8}{63} l_\Lambda^2\ .
\eeq
Notice that in the former sections \ref{sec:quad} and \ref{sec:displacement} we had already derived all the parameters  $l_S,\,l_{T},\,l_{TF},l_{\Phi_S}$ that were needed to regularize not only $\vec s$, but also the quadrupole of Lagrangian regions. We can easily check that, by using the values for $l_S,\,l_{T},\,l_{TF},l_{\Phi_S}$ found in the former subsection, we find the same number for $l^{\rm c.t.}_{s, \rm comb}$:
\bea
&&\frac{1}{36}\left( l_T(\Lambda,\eta_0)^2 + 2 l_{TF}(\Lambda,\eta_0)^2 +2l_S(\Lambda,\eta_0)^2+ 6l_{\Phi_S}(\Lambda,\eta_0)^2\right)_{\rm c.t.}= \\ \nn
&&\qquad \frac{1}{36}\left(-2-\frac{4}{7} + \frac{8}{35} -6\times \frac{121}{105} \right) l_\Lambda^2 =\frac{8}{63} l_\Lambda^2\ ,
\eea
as expected. Therefore, as long as we are interested in correlations of the displacement, we only need one parameter to make predictions at one-loop: $l_{s,\rm comb}$, which can be derived directly by matching to observations of the displacement field, or alternatively by obtaining $l_S,\,l_{T},\,l_{TF},l_{\Phi_S}$ independently, which carry additional information.

If we are furthermore interested in correlations of the matter density, $\delta_m$, which is defined through~\eeqref{equation2}: $\delta_m \propto \partial_x^2\Phi$, it is clear that we encounter a new combined parameter. In fact, as it is well known,  from \eeqref{eq:delta_LPT_fourier}, the matter density correlation function and the power spectrum can be written as 
\bea
\label{densityre}
&&1+\langle\delta_{m,L}(\vec x_1,\eta_1)\,\delta_{m,L}(\vec x_2,\eta_2)\rangle = \int d^3 q~ \int_k e^{-i\vec k\cdot (\vec q - \vec r)} \left\langle e^{-i\, \vec k\cdot (\vec s(\vec q,\eta_2)-\vec s(\vec 0,\eta_1))}\right\rangle\ ,\\ \nonumber
&&\quad\Rightarrow \quad  \langle\delta_{m,L}(k,\eta_1)\delta_{m,L}(k,\eta_2)\rangle'=\int d^3 q\; e^{-\vec k\cdot \vec q}  \left\langle e^{-i\, \vec k\cdot (\vec s(\vec q,\eta_2)-\vec s(\vec 0,\eta_1))}\right\rangle\ ,
\eea
with $\vec q = \vec q_2 - \vec q_1$ and $\vec r=|\vec x_2-\vec x_1|$. Due to the homogeneity, the average only depends on the separation. On the other hand, using \eqref{dmkspa}, in LEFT the above equation is modified to
\bea\label{eq:LEFT_corre} 
&&1+\langle\delta_{m,L}(\vec x_1,\eta_1)\,\delta_{m,L}(\vec x_2,\eta_2)\rangle =\\ \nn
&&\qquad\qquad \int d^3 q~ \int_k e^{-i\,\vec k\cdot (\vec q - \vec r)} \left\langle e^{-i\, \vec k\cdot (\vec s_L(\vec q,\eta_2)-\vec s_L(\vec 0,\eta_1))-\frac{1}{2}k^i k^j (Q^c_{ij}(\vec q_1,\eta_1) + Q^c_{ij}(\vec q_2,\eta_2))+\ldots}\right\rangle\ , \\ \nonumber
&&\quad\Rightarrow \quad  \langle\delta_{m,L}(k,\eta_1)\delta_{m,L}(k,\eta_2)\rangle'=\int d^3 q\; e^{-\vec k\cdot \vec q}  \left\langle e^{-i\, \vec k\cdot (\vec s(\vec q,\eta_2)-\vec s(\vec 0,\eta_1))-\frac{1}{2}k^i k^j (Q^c_{ij}(\vec q_1,\eta_1) + Q^c_{ij}(\vec q_2,\eta_2))+\ldots}\right\rangle\, .
\eea
By Taylor expanding the exponential in $\vec s$, straightforward algebra shows that in the equal-time matter power spectrum the counter-terms combine as
\be
P_{\delta_m\delta_m}(k)\supset - \frac{1}{3}l_{\delta_m,\rm comb}(\Lambda,\eta_0){}^2 k^2 P_L(k)
\ee
with
\beq
l_{\delta_m,\rm comb}(\Lambda,\eta_0){}^2 \equiv l_{s,\rm comb}(\Lambda,\eta_0){}^2 +  l_S(\Lambda,\eta_0)^2 +l_T(\Lambda,\eta_0)^2+2 l_{TF}(\Lambda,\eta_0)^2\ .
\eeq
This  could have been equally derived by inspection of \eeqref{equation2}, keeping in mind that the counter-terms need to be evaluated at linear level. Using the definition of $l_{\delta_m,\rm comb}$ above one can easily show the divergences in $\langle \delta_m \delta_m\rangle$ are also regularized by our previous choices.

Notice that in the limit that all correlation functions are evaluated after the fluctuating fields and counter-terms are brought down from the exponential, there is no difference between LEFT and the Eulerian counterpart. This means that, working in this approximation, one can simply replace $l_{\delta_m,\rm comb}(\Lambda,\eta_0)$ with the analogous parameter called $c_{s,\rm comb}$
 in the Eulerian EFT~\cite{Carrasco:2012cv,Carrasco:2013mua}, without the need of a new independent fit. The relation between the two is simply found by matching the form of the power spectra~\footnote{For example in the notation of ~\cite{Carrasco:2013mua}:
\be\label{eq:l_comb_delta}
l_{\delta_m,\rm comb}(\Lambda,\eta_0){}^2= 2\pi\; c_{s}^2\cdot \frac{1}{k_{\rm NL}^2}\ .
\ee
Furthermore $c_{s}^2$ is sometimes referred to as $c_{\rm counter}$~. Both $c_{s}$ and $c_{\rm counter}$ are dimensionless.}.

 Of course, one might wonder if one can compute the correlation functions in \eeqref{eq:LEFT_corre} {\it without} Taylor expanding in $\vec s$. It is in this regime that LEFT becomes extremely useful. We discuss this next.
 
\section{Resummed correlation function}\label{resum}

As we mentioned above, in LPT the matter density correlation function in real space can be written as~\cite{Matsubara2}:
\beq
\label{densityre2}
1+\xi(\vec r) = \int d^3 q~ \int_ke^{-i\vec k\cdot (\vec q - \vec r)} \left\langle e^{i\, \vec k\cdot (\vec s(\vec q,\eta)-\vec s(\vec 0,\eta))}\right\rangle\ ,
\eeq
where we defined the equal time matter correlation function as
\be
\xi(\vec r)=\langle\delta_{m,L}(\vec x_1,\eta_1)\,\delta_{m,L}(\vec x_2,\eta_2)\rangle\ ,
\ee
with $\vec r=\vec x_2-\vec x_1$. This expression can be evaluated as follows (following the notation in \cite{Carlson:2012bu}):
\bea
\label{kcumua}
1+\xi(\vec r) &=&  \int d^3 q~ \int_k e^{-i\vec k\cdot (\vec q - \vec r)} K(\vec q,\vec k) \ , \nonumber \\
K(\vec q,\vec k)&\equiv& \langle e^{i\, \vec k \cdot \Delta \vec s}\rangle  = {\rm exp}\left[\sum_n\frac{i^n}{n!} k_{i_1} \ldots  k_{i_n} \langle  \Delta s^{i_1} \ldots \Delta s^{i_n}\rangle_c\right]\ ,
\eea
where $\Delta \vec s \equiv \vec s(\vec q,\eta)-\vec s(\vec 0,\eta)$, and $\langle  \Delta s^{i_1} \ldots \Delta s^{i_n}\rangle_c$ are the connected moments or cummulants. Given that we are interested in a one-loop calculation we only need up to the third moment. Hence, \eqref{kcumua} becomes
\beq
\label{logk}
\log K = -\frac{1}{2} A_{ij}(\vec q) k^i k^j - \frac{i}{6} W_{ijl}(\vec q) k^i k^j k^l +  \ldots\ ,
\eeq
where
\bea
A^{ij}(\vec q) &\equiv& \langle \Delta s^i \Delta s^j \rangle_c\ , \\
W^{ijl}(\vec q) &\equiv& \langle \Delta s^i \Delta s^j \Delta s^l \rangle_c\ .
\eea


These expressions involve products of fields at the same point and thus can lead to divergencies. Moreover in general they need to be renormalized even if they are finite. In LEFT, the expression for the correlation function reads: 
\bea\label{kcumuleft}
1+\xi_L(\vec r) &=& \int d^3 q~ \int_ke^{-i\,\vec k\cdot (\vec q - \vec r)} \left\langle e^{i\, \vec k\cdot (\vec s_L(\vec q,\eta)-\vec s_L(\vec 0,\eta))-\frac{1}{2}k^i k^j (Q^c_{ij}(\vec q_1,\eta) + Q^c_{ij}(\vec q_2,\eta))+\ldots}\right\rangle\ , \nonumber \\
 &=&  \int d^3 q~ \int_k e^{-i\vec k\cdot (\vec q - \vec r)} \,K_L(\vec q,\vec k)\ .
 \eea
Keeping up to the third moment, $K_L$ reads: 
\beq
\label{logkL}
\log K_L = -\frac{1}{2} A_L^{ij}(\vec q) k_i k_j - \frac{i}{6} W_L^{ijl}(\vec q) k_i k_j k_l +  \ldots\ ,
\eeq
where
\bea
A_L^{ij}(\vec q) &\equiv& \langle \Delta s^i \Delta s^j \rangle_c + 2 \langle Q^{ij}\rangle_S\ , \\
W_L^{ijl}(\vec q) &\equiv& \langle \Delta s^i \Delta s^j \Delta s^l \rangle_c + \Delta \langle  s^i Q^{jk}_{\cal R} \rangle + \Delta \langle s^j Q^{ik}_{\cal R} \rangle + \Delta \langle  s^k Q^{ij}_{\cal R} \rangle\ ,
\eea
where $ \Delta \langle  s^i Q^{jk}_{\cal R} \rangle  = \langle  s^i(q) Q^{jk}_{\cal R}(0) \rangle -  \langle  s^i(0) Q^{jk}_{\cal R}(q) \rangle$.
By construction the one-loop order  divergences that appear in these expressions have been regularized because they involve either  the two point function of the displacement or correlations involving the quadrupole that we regularized using $l_{TF}^2,\,l_{T}^2, l_S^2$ in section \ref{sec:oneloop}. This is exactly why we regularized the quadrupole  in an explicit way.

The correlations between composite operators  involving powers of the displacement at the same point are regularized and renormalized by the multipole moments which appear in the expression for the density in LEFT. One could also take the point of view that the dynamical equations, the formula for the density, etc., are left as in LPT but each composite operator that appears needs to be properly renormalized, a fact that introduces new renormalization parameters. We think discussing LEFT as describing the dynamics of extended objects is more intuitive but both points of view are equivalent.

Formulas such as (\ref{kcumuleft}) depend on the exponential of connected moments. At least naively these formulas should not be trusted in the sense that by keeping those terms in the exponent one is keeping contributions that should be subleading relative to terms that have been dropped. In our case for example we stopped at the third moment but higher powers of these contributions coming from the exponentiation we are implicitly keeping are subleading with respect to the forth order and higher moments we are neglecting.

Two comments are in order. Of course, there is no problem in keeping subleading terms as long as they are not anomalously large, and as long as one is aware of it and properly quantifies the uncertainty in the calculation by estimating the size of the terms that have been neglected. In this case, the subleading contributions being kept would be small compared to this theoretical error bars and thus would not be trusted automatically. One may be keeping the terms for computational simplicity but only trusting the result where appropriate.

There is however a sense in which the terms being kept do add additional information and this is what we want to discuss next. This is related to the fact that there is more than one parameter that sets the relative sizes of the various terms in the perturbative series. Keeping terms in the exponential consistently sums some of the biggest corrections, and thus it provides a better estimate of the correlation function, even after appropriately estimating the errors from the terms that are being missed. Even more generally, it is justified to sum up some contributions when they are parametrical distinguished from the others. Of course, this is particularly useful when those contributions are the larger ones. This situation is most striking for non-power law initial spectra, and further enhanced in our universe by the presence of the BAO peak in the correlation function. Thus, in a sense, this discussion is outside the scope of this paper. We include it because it provided the motivation for us to develop the EFT in Lagrangian space in the first place. We will present the application of LEFT to our universe in a separate paper, where the discussion summarized here will have more concrete consequences for the results.   
 
\subsection{Expansion parameters of perturbation theory} 

The first point in the argument is to explicitly remind the reader that there are terms of various different sizes at a fixed order in SPT (See Fig. \ref{epsilons}).  It is pedagogically more transparent to show the results of SPT in real space. We will follow the notation of \cite{Sherwin:2012nh}. 

The standard perturbative solution for $\delta$ reads: 
\bea
\dl(\bx) &=& \dl^{(1)}(\vec x)+ \dl^{(2)}(\vec x) + \dl^{(3)}(\vec x) + \cdots \ ,
\eea
where up to the second order term is given by:
\bea\label{delta2}
\dl^{(2)}(\bx)&=&d_i^{(1)} \partial_i \dl^{(1)}+ {17\over 21} \left(\dl^{(1)}\right)^2 + \frac{2}{7} K_{ij}^{(1)} K_{ij}^{(1)}\ , \nonumber \\
d_i^{(1)}(\bx)& =& - \int {d^3 k \over (2\pi)^3} {i k_i \over k^2} \, \dl^{(1)}(\vec k)\, e^{i \vec k\cdot\vec x} \nonumber\ , \\
K_{ij}^{(1)}(\bx)& =&  \int {d^3 k \over (2\pi)^3} \left({ k_i k_j\over k^2}-{1 \over 3}\delta_{ij}\right)  \dl^{(1)}(\vec k) \, e^{i \vec k\cdot\vec x}\ .
\eea
We are interested in understanding the effect of a long wavelength mode on a short fluctuation. We can use this equation to notice an important point: there are two types of terms in this expression which could have very different sizes. The second two terms in the expression of $\delta^{(2)}$ are down with respect to the linear field by a factor of $\delta^{(1)}$. The first term is quite different, as it involved the displacement produced by the long mode and a gradient acting on the short mode. If this first term is the dominant one, related higher order terms can be resummed easily. 

The expression for $\delta^{(3)}$ is complicated and given in the appendix of \cite{Sherwin:2012nh}. The important point here is that if one neglects effects proportional to the density contrast produced by the long mode then one simply has $\delta^{(3)} = \tfrac{1}{2}\, d_i^{(1)}  d_j^{(1)}\, \partial_i\partial_j \dl^{(1)} + \cdots\,$. 
In other words, if we keep the terms  from the displacement only, we obtain:
\be
\label{shifted}
\delta(\bx)=\delta^{(1)}+d_k^{(1)} \partial_k \delta^{(1)}+ {1\over 2} d_k^{(1)} d_l^{(1)} \partial_k\partial_l\delta^{(1)} + \cdots = \delta^{(1)}(\vec x +  \vec{d} )\ .
\ee
It is this shift that is responsible for the majority of the smearing seen in the BAO peak of the correlation function, and thus, when properly modeled, leads to remarkable agreement between model and simulation (see for example \cite{Sherwin:2012nh}). Although the analysis of the BAO feature is beyond the scope of this paper, we present a simple toy model to illustrate our point in Appendix~\ref{toymodel}.

We can connect with the previous sections of the paper by noticing that the relation between our Lagrangian displacements and the number density field is \eeqref{mapqx}, that we repeat here for convenince: 
\beqa
\label{mapqx2}
1+\delta({\vec x},\eta)&=& \int  d^3{\vec q}~\delta^3( {\vec x} - {\vec z}({\vec q},\eta)) \nonumber \\
&=& \left[{\rm det} \left(\tfrac{\partial z^i}{\partial q^j}\right)\right]^{-1} =\left[{\rm det} \left(1+\tfrac{\partial s^i}{\partial q^j}\right)\right]^{-1},
\eeqa
where the second line is evaluated at the solution of ${\vec z}({\vec q},\eta)=\vec x$, or equivalently ${\vec q}= \vec x - \vec s(\vec q)$. 
There are various non linear terms in this equation: one can expand the determinant to higher powers in ${\partial s^i}/{\partial q^j}$, one can consider higher order solutions of $\vec{s}$, terms like ${\partial s^{(2)i}}/{\partial q^j}$, and finally one has the effects from the coordinate transformation
${\partial s^i}/{\partial [q^j}(\vec x -\vec s)] =  {\partial s^i}/{\partial q^j} + s^l {\partial^2 s^i}/{\partial q^j\partial q^l} + \cdots$.  
These last ones are the terms we are focusing on in this section. 

\subsection{Higher order terms in the displacement}

The terms that involve only higher order powers of the displacement are easy to get in general, so if these are the most relevant ones, one can include higher powers of them with full theoretical control.  Let us imagine that, for some range of modes, we only care about the terms that  involve the displacement they produce, rather than terms proportional to their density contrast. In this case, we are considering a gravitational potential that is approximately linear over the region of interest, as we are neglecting the second derivatives of the potential produced by those modes (which is proportional to the density). As a result of the equivalence principle, a linear potential can be removed by a coordinate transformation (see for example~\cite{Carrasco:2013sva} where this fact is used to explain why the equal-time matter power spectrum is unaffected by IR modes if $n>-3$; and \cite{Creminelli:2013mca} (see also \cite{Riotto,Peloso:2013zw,Creminelli:2013poa,Kehagias:2013rpa}) where this fact is used to obtain ``consistency conditions" for higher order correlation functions in LSS). This means that, to this level of accuracy, the modes only enter through the coordinate transformation in equation (\ref{shifted}), and not in the higher order terms such as ${\partial s^{(2)i}}/{\partial q^j}$. The coordinate transformation is automatic in Lagrangian space, the only thing we need to do is add the displacement $\vec s$ produced by the relevant modes. 

If we want this approximation to be useful, it needs to be the case that the displacement is dominated by large scales, contrary to what happens for the density, which is dominated by the small scales for $n>-3$. Under these conditions, one can be in a regime in which a large part of the displacement is produced by modes that induce very small density fluctuations and tides, and thus have small dynamical effects. This is the case for our universe, an in general if $n<-1$. This also makes  the displacement UV finite. Note that for power law initial conditions with $n<-1$, the displacement is IR divergent, and thus results in large corrections in unequal time correlation functions. Being able to understand and resum their contributions seems therefore crucial.

In our universe, the presence of the BAO peak in the correlation function also makes these terms even more important. In fact, if we want to compare the effect of the displacement to the density, we need to remind ourselves  that, if we keep a displacement, then the term must have a derivative acting on the correlation function, as each displacement always comes with a derivative (see for example equation (\ref{delta2})).  So, for example, to asses the relevance of the terms with the displacement relative to those containing higher powers of the density, we need to compare  $\delta_L^2\, \xi$ with $d_L^2\, \nabla^2 \xi$, where $\delta_L$ and $d_L$ are the density contrast and displacement produced by the long mode. In our universe, because of the BAO and the shape of the power spectrum, the displacement term is much larger. The toy model in appendix~\ref{toymodel} makes this point in more detail.  Thus in our universe it makes sense to try to keep more terms related to the change of coordinates.

If  we neglect the density contrast produced by the long modes, they do not affect the moments of the displacement produced by the short modes. For example by treating the long modes as Gaussian variables and keeping them in the exponential, we are including correctly all the higher order terms involving only the displacement, but we are not treating consistently higher order terms involving the density contrast they produce. Thus, the only benefit of keeping terms in the exponential is that we are adding all the terms with higher powers of the displacement consistently, but we are not gaining anything from higher order terms involving more powers of the density. Those terms can be either brought down or kept in the exponential, provided we are aware that their associated higher order contributions are of the  same size of terms neglected, and thus are smaller than the theoretical error bars. We are also making a mistake in the size of the displacements if we just compute them at some fixed order, say linear order. To the extent the modes that dominate the displacement are large modes, their non-linear corrections are small. But this needs to be taken into account when estimating the errors.

Let us add a final comment on the number of unknown coefficients that are needed at a given order. As we showed in section~\ref{sec:comb}, if we put all terms in the exponents in \eeqref{kcumuleft} downstairs, all the unknown parameters associated to the multipoles combine so that, in the one-loop calculation of the matter power spectrum, only one unknown parameter is necessary to renormalize the theory: the $l_{\delta_m,\rm comb}$ of \eeqref{eq:l_comb_delta}. If instead we were to keep the multipoles upstairs in the exponential in \eeqref{kcumuleft}, then the parameters would not combine in one, and we would need several parameters to renormalize the one-loop calculation. This is a consistent conclusion. However, we have seen in this discussion, the point of keeping the terms in the exponent is to resum the contribution from the displacements due to infrared modes. The role of the counter-terms is instead to consistently include the effect of the short distance physics, for which keeping terms upstairs in the exponential is irrelevant. A mixed approach therefore emerges. For a scaling universe, one could keep in the exponent only the terms that involve the counter-terms and that have the role of cancelling the divergent parts, so that the result is cutoff independent. One could instead bring downstairs the terms that involve the finite coefficients of the multipoles. At this point the finite terms will combine into only  one coefficient. For the true universe,  as we will discuss explicitly in a subsequent paper, a similar approach is expected to work.

In summary: working in Lagrangian space makes adding higher order terms involving the displacement straightforward.  We are only keeping terms in the exponential to sum these displacement terms.  This is only useful for large modes whose dynamical effects are small. Thus the displacement needs to be dominated by the IR modes. Furthermore, it is the presence of the BAO peak that makes this even more useful, enhancing the size of the derivatives of the correlation function.

\section{Effective action approach}\label{sec:action}

In this section we will re-derive our previous results using an effective action approach. This will allow us to set up a formalism which can be readily extended to go beyond the Newtonian approximation, as well as to simply include other types of matter and interactions, such as baryons. It also allows, in principle, to quantize  the system in a simple way, something that is hardly of relevance in our context. Finally, having at our disposal an action, some of the procedures for regularizing and renormalizing the correlation functions may become simpler. The formalism used in this section bears a close resemblance with the approach introduced to study gravitationally bound extended objects \cite{nrgr1,nrgr2,nrgr3,disip1,disip2}, albeit in the continuum limit and in a non-relativistic setting.

\subsection{Dark matter point particles \& Newtonian limit}

As before we consider a set of point-like dark matter particles interacting gravitationally in an FLRW background, with the universe dominated by dark matter and a cosmological constant. The action reads
\be
\label{stot}
S_{\rm tot} =\int d^3{\vec x}d\eta\sqrt{-g}\;\left[-\frac{1}{16\pi G}R+\Lambda \right]+S_{\rm pp}^{\rm DM},
\ee
with $S_{\rm pp}^{\rm DM}$ given by
\beq
S_{\rm pp}^{\rm DM}= -\sum_A m_A \int d^3{\vec x} d\eta_A ~\delta^3({\vec x}-{\vec z}_A(\eta_A)) \sqrt{-g_{\mu\nu}({\vec x},\eta)\dot z^\mu_A(\eta_A)  \dot z^\nu_A(\eta_A)} \ .
\eeq
In this expression $z^\mu_A(\eta_A)$ represents the worldline co-moving coordinates for the $A$-particle, and dots are taken with respect to $\eta$. In what follows, since we are interested in the dynamics around an FLRW background, we will set coordinates where $z^0\equiv \eta(t)$, and $\eta_A = \eta$ for all particles.

We now to take the continuum limit over the index $a$, i.e. 
\be
\sum_A m_A \to \int  \tilde\rho_{in}({\vec q})\;a_{in}^3 ~d^3{\vec q}\ .
\ee  
$\tilde\rho_{in}({\vec q})$ represents the mass density per unit $d^3{\vec q}$ cell, at some initial time $\eta_{in}$, when the scale factor is $a(\eta_{in})=a_{in}$. By properly choosing the initial displacement ${\vec z}({\vec q},\eta_{in})$, we can always take 
\be
\label{rhoq}
\tilde\rho_{in}({\vec q})=\bar\rho_m(\eta_{in})=\left(\frac{3}{8\pi G}  { H}^2 \Omega_m\right)_{\eta=\eta_{in}},
\ee
namely, we can take $\tilde\rho_{in}({\vec q})$ to be equal to  the homogeneous dark matter component at time $\eta_{in}$. We will take the limit of $\eta_{in} \to 0$. Hence we re-write this action as
\bea
&&S_{\rm pp}^{\rm DM}=  -\int d^3{\vec x} d^3{\vec q} d\eta\;\bar\rho_m(\eta_{in})\;a_{in}^3 \; \delta^3({\vec x}-{\vec z}({\vec q},\eta))\; \sqrt{-g_{\mu\nu}({\vec x},\eta)\dot z^\mu({\vec q},\eta)  \dot z^\nu({\vec q},\eta)} \nn \\ 
&&\qquad=-\int d^3{\vec x} d^3{\vec q} d\eta\;\bar\rho_m(\eta)\;a(\eta)^3  \delta^3({\vec x}-{\vec z}({\vec q},\eta))\; \sqrt{-g_{\mu\nu}({\vec x},\eta)\dot z^\mu({\vec q},\eta)  \dot z^\nu({\vec q},\eta)}, \label{actiondm}
\eea
where in the second passage we have used how the background density redshifts as $\bar\rho_m(\eta_{in})\;a^3(\eta_{in})=\bar\rho_m(\eta)\;a^3(\eta)$. The unperturbed solution is given by a static map (in co-moving coordinates) $\bar { z}^i= {q}^i$, which corresponds to the FLRW background 
\beq
\bar g_{\mu \nu} dx^\mu d x^\nu = a^2(\eta)\left(-d\eta^2 + d{\vec x}^2\right).
\eeq  
We are interested on the dynamics of the dark matter particles on scales much shorter then Hubble, where the non-relativistic approximation is well justified. Therefore consider the FLRW Newtonian gauge,
\be
ds^2=a^2\left[-(1+2\Phi) d\eta^2+(1-2\Psi)dx^2\right]\ ,
\ee
to leading order in $\Phi, \Psi$. Moreover, we will work in the Newtonian approximation: $\tfrac{1}{a(\eta)}\pd_\eta { z^i(q,\eta)}\ll 1$ and $   \pd_i \gg  \pd_t \sim H$.  In this limit the matter action (\ref{actiondm}) becomes
\bea
\label{action2}
&&S_{\rm pp}^{\rm DM}=  \int d^3{\vec x} d^3{\vec q} d\eta\;\bar\rho_m(\eta)\;a^4 \; \delta^3({\vec x}-{\vec z}({\vec q},\eta))\; \left(-1+\frac{1}{2}\left(\frac{d{z^i}({\vec q},\eta)}{d\eta}\right)^2-\Phi({\vec x},\eta)\right) \ .
\eea
The terms in the action that are linear in the metric fluctuations lead to the background equations, whose solutions is
\be
H^2=\frac{8\pi G}{3}\left[\Lambda+\bar \rho_{m}(\eta)\right]\ , \qquad \frac{d H}{dt }=-\frac{8\pi G}{3}\bar\rho_{m}(\eta).
\ee

The above results suggest the following manipulations. Let us first introduce the mass density per unit $q$-cell:
\bea
\tilde\rho({\vec z}({\vec q},\eta),{\vec x})&\equiv&\bar\rho_{m}(\eta) ~\delta^3({\vec x}-{\vec z}({\vec q},\eta))\ ,\\
\delta\tilde\rho({\vec z}({\vec q},\eta),{\vec x}) &\equiv& \tilde\rho({\vec z}({\vec q},\eta),{\vec x}) -\bar\rho_{m}(\eta)\delta^3({\vec x}-{\vec q})=\bar\rho_m(\eta)\left[\delta^3({\vec x}-{\vec z}({\vec q},\eta))-\delta^3({\vec x}-{\vec q})\right]\ .
\eea
We can then add and subtract to the action the term
\be
\int d^3{\vec x}d\eta \;a(\eta)^4\, \bar\rho_{m}(\eta) \Phi({\vec x},\eta)=\int d^3{\vec x}d^3{\vec q}d\eta\; a(\eta)^4\, \bar\rho_{m}(\eta) \delta^3({\vec x}-{\vec q}) \Phi({\vec x},\eta)\ .
\ee 
One of them will cancel the tadpole contribution in the Einstein-Hilbert action, while the other will combine with $\rho({\vec z}({\vec q},\eta),{\vec x}) \Phi({\vec x},\eta)$ to form $\delta\rho({\vec z}({\vec q},\eta),{\vec x}) \Phi({\vec x},\eta)$.  We are thus finally led to
\beq
\label{action3}
S_{\rm tot}= \int a^4 d \eta \int d^3{\vec x} d^3{\vec q} \left\{\frac{1}{2}\, \tilde\rho ({\vec z}({\vec q},\eta),{\vec x})\left[-1+\left(\frac{d{\vec z}({\vec q},\eta)}{d\eta}\right)^2\right]  -\delta\tilde\rho({\vec z}({\vec q},\eta),{\vec x})\, \Phi({\vec x},\eta)\right\} + S_{EH}^{(2)}\ .
\eeq
The term $S_{EH}^{(2)}$ is the quadratic part of Einstein-Hilbert action. In the Newtonian limit, after solving the constraint equation that gives $\Psi= \Phi$ and plugging this back into the action, it takes the form
\be\label{EHnewt}
S_{EH}^{(2)}=-\frac{1}{8\pi G}\int d^3x d\eta\; a^4 \frac{1}{a^2}\pd_i\Phi\pd_i\Phi \ .
\ee
We have now managed to write an action directly for the fluctuations, which, as it should be, has no tadpole terms.

Varying the action as usual we obtain Einstein's equation, 
\be
\ \frac{\pd^2}{a^2}\Phi=4\pi G \int d^3{\vec q}\;\delta\tilde\rho({\vec z}({\vec q},\eta),{\vec x})\ ,
\ee
which can be written as
\be
\pd^2\Phi(x)=4\pi G a^2~\delta\rho(x)=\frac{3}{2}{\cal H}^2(\eta)\Omega_m(\eta) \delta(x)\ ,
\ee
where, as usual, ${\cal H}=a H$, and where we have defined
\be
\delta\rho(x)\equiv\int d^3{\vec q}\;\delta\tilde\rho({\vec z}({\vec q},\eta),{\vec x})\ ,\quad \delta({\vec x},\eta)=\frac{\delta\rho({\vec x},\eta)}{\bar\rho_m(\eta)}\ .
\ee
On the other hand, for the equation of motion of the particle's trajectory, we have
\be
\int d^3 x\left[ \frac{d}{d\eta}\left(a^4 \bar\rho_{m}(\eta)  \delta^3({\vec x}-{\vec z}({\vec q},\eta))  \frac{d{ z^i}({\vec q},\eta)}{d\eta} \right)+ a^4 \bar\rho_{m}(\eta)  \delta^3({\vec x}-{\vec z}({\vec q},\eta))  \frac{\pd\Phi}{\pd x^i}\right]=0\ ,
\ee
where we used that $\pd_{\vec q}\delta^3({\vec x}-{\vec q})=-\pd_{\vec x}\delta^3({\vec x}-{\vec q})$, and we have integrated the $x$-derivative by parts.  Performing the $d^3x$-integral, we finally obtain
\be
\frac{d^2 {z^i}({\vec q},\eta)}{d \eta^2} +{\cal H}\frac{d{ z^i}({\vec q},\eta)}{d\eta} =  -\left. \pd_i \Phi\right|_{{\vec x}={\vec z}({\vec q},\eta)}\ 
\ee

These equations are the traditional ones obtained in the literature, now derived from an action principle. Next we construct an effective action where we separate the short and long distance modes in \eqref{stot} in a derivative expansion.
\subsection{The effective action for the long-distance universe}

In this section we construct an EFT description of the long-distance universe after integrating out the short modes. Using the symmetries of the problem we can write an effective action where the dynamics of the short distance degrees of freedom is integrated out. This action will be a generalization of \eqref{actiondm}, and takes the form
\bea
\label{effact1}
 S_{\rm LEFT} &=& S_{L}^{\rm DM} - \int a^3 d\eta \int d^3{\vec q} d^3{\vec x}~ \tilde\rho_L({\vec z}_L({\vec q},\eta),{\vec x}) \left\{\frac{1}{2} {\dot z}_L^\mu \omega_\mu^{ab}(\vec x,\eta) L^{ab}({\vec q},\eta) + \tfrac{1}{2}Q^{ab}_{E}({\vec q},\eta) E_{ab}(\vec x,\eta)  \right.  \nn \\ &+& \tfrac{1}{2} J^{ab}_{B}({\vec q},\eta) B_{ab}(\vec x,\eta)+ \frac{1}{6} Q^{abc}(\vec q,\eta) \nabla_c E_{ab}(\vec x,\eta) + \frac{1}{2} J^{abc}(\vec q,\eta) \nabla_c B_{ab}(\vec x,\eta) + \ldots \nn\\
 &+& \left. P^a({\vec q},\eta) R_{a\mu}(\vec x,\eta) {\dot z}_L^\mu(\vec q,\eta)+C({\vec q},\eta) R(\vec x,\eta) + C_V({\vec q},\eta) R_{\mu\nu}(\vec x,\eta){\dot z}_L^\mu(\vec q,\eta) {\dot z}_L^\nu(\vec q,\eta) +\ldots \right\}\ , \nn \\ 
\eea
where we defined the  {\it monopole} term as given by
\beq
S_L^{\rm DM} \qquad=-\int d^3{\vec x} d^3{\vec q} d\eta\;\rho_E(\vec z_L(\vec q,\eta),\vec x)\;a^3 \; \sqrt{-(\eta_{\mu\nu}+h^L_{\mu\nu}({\vec x},\eta))\dot z_L^\mu({\vec q},\eta)  \dot z_L^\nu({\vec q},\eta)}.
\eeq
This is the same expression as in \eqref{action2} with $\vec z \to \vec z_L$, $h_{\mu\nu} \to h_{\mu\nu}^L$, and  
\beq
\rho_E(\vec z_L(\vec q,\eta),\vec x) \equiv \bar\rho_m( \eta) \left[ 1 + V_S(\vec q, \vec x,\eta)\right] \delta^3 (\vec x - \vec z_L(\vec q,\eta)),
\eeq
where $\rho_E(\vec z_L(\vec q,\eta),\vec x)$  is the energy density associated with the extended object labelled by $\vec q$. It includes not only the contribution from the total mass of the particles but also a new term we denote $V_S(\vec q, \vec x,\eta)$. In the example studied in this paper the new term results from gravitational interactions with particles in other Lagrangian regions that overlap with the finite sized object at $\vec z_L(\vec q,\eta)$~\footnote{If we were interested in relativistic corrections,  $\rho_E$ should also include contributions from the internal potential and kinetic energies of the particles. However, these pieces are not relevant at the order we are working.}. Note $V_S(\vec q,\vec x,\eta)$ depends on the distribution of matter inside the particle and that is why it carries an argument $\vec q$, but also on the internal distribution of the overlapping regions which have position $\vec x$. This term is evaluated at the center of mass of the extended object: $\vec x=\vec z_L(\vec q,\eta)$. Detailed derivations of the expression for $V_S(\vec q, \vec x,\eta)$ are presented in \ref{uvmatchone} and \ref{matching}. From the point of view of the long wavelength EFT we will need to parametrize the response of $V_S(\vec q, \vec x,\eta)$ to a long wavelength mode.


For the other terms in \eeqref{effact1}, we have
\bea
\tilde\rho_L({\vec z}({\vec q},\eta),{\vec x})&\equiv& \bar\rho_m(\eta) ~\delta^3({\vec x}-{\vec z}_L({\vec q},\eta))\ ,\\
\delta\tilde\rho_L({\vec z}({\vec q},\eta),{\vec x}) &\equiv& \tilde\rho_L({\vec z}({\vec q},\eta),{\vec x}) -\bar\rho_m(\eta)\delta^3({\vec x}-{\vec q})\\ &=&\bar\rho_m(\eta)\left[\delta^3({\vec x}-{\vec z}_L({\vec q},\eta))-\delta^3({\vec x}-{\vec q})\right]\nn\ .
\eea
Finally, the ellipses account for higher order terms in derivatives and powers of $h^L_{\mu\nu}$. Here $z^\mu_L=(\eta,{\vec z}_L)$ is the position of the smoothed mass density whose dynamics we are following in the long distance universe: it represents the center-of-mass of a large fraction of the dark matter particles which make up each region where we integrate out the short distance dynamics.

In the above expression \eeqref{effact1}, quantities with the $\vec q$-argument represents (extended) bodies and their moments, while quantities with the $\vec x$-arguments represent fields that they interact with. In particular, $E_{ab}, B_{ab}$ are the electric and magnetic components of the Weyl tensor in a locally flat frame parallel transported by the center-of-mass defined by $e^\mu_a$ (with $e^\mu_0 = \dot z_L^\mu$), $L_{ab}$ is the angular momentum, $R_{\mu\nu}$ is the Ricci tensor, and $\omega^{ab}_\mu$ are the Ricci rotation coefficients. The extra (higher dimensional) terms are multiplied by a set of time dependent (space-like) multipole moments which are obtained through a matching computation. This construction will become more transparent when we perform this matching explicitly later on.
In \eqref{effact1} only the long distance degrees of freedom are kept in the EFT. Notice our effective action is invariant under full diffeomorphisms.

Several assumptions went into the particular form of (\ref{effact1}), that ultimately come from the UV model that we have in mind, which we described in the previous section. In our context, we are mainly interested in the Newtonian description of dark matter particles that interact only gravitationally. This constraints the type of operators that we used in (\ref{effact1}), on top of what allowed by purely symmetry reasons. Firstly, since the UV action in \eqref{actiondm} is linear in the Newtonian potential, the effective theory must also be linear, which in principle restricts terms quadratic in $h^L_{\mu\nu}$~\footnote{Obviously the terms already present in \eqref{effact1} do induce non-linear effects, through their response functions, but nonetheless are the complete set that encodes the Newtonian limit.}. Secondly, the fact that particles interact only gravitationally also forbids the presence of multi-particle vertices, such as terms that would appear in the Lagrangian in the form of multiple $\vec q$-integrals.
If the interactions are short distance with respect to the scale of validity of the EFT, these terms reduce to contributions that include derivatives of $\vec z(\vec q,\eta)$ with respect to $\vec q$, for example 
 \be
 \label{contact}
 \int d\eta \int d^3\vec q\;  {\cal L}(\pd_{q^j}{z}^i\pd_{q^j}{z}^i,\; \pd_{q^iq^l}{z}^j\pd_{q^i q^l}{z}^j,\ldots)\ .
 \ee
These interactions could be straightforwardly included if one wished to describe at low energies different UV models, for example if one includes baryons. We will comment on this possibility later on~\footnote{Notice that, even though our short modes have a time-scale comparable to the one of the long modes, the effective action is still local in time. The relatively-long time-scale of short modes enters in making the response of the multipole non-local in time. 
}.

If we now ignore all relativistic effects, which are sub-leading in $v/c \ll 1$, such as rotational degrees of freedom, and perform a similar expansion as in \eqref{action2}, then \eqref{effact1} becomes (see appendix~\ref{matching} for details)
\bea
&& S_{\rm LEFT} = \int d \eta \int d^3{\vec x} d^3{\vec q}\;a(\eta)^4  \label{effnewt}\\ \nn
&&  \qquad\qquad\qquad \left\{-\rho_E({\vec z}_L({\vec q},\eta),{\vec x})+\tilde\rho_L({\vec z}_L({\vec q},\eta),{\vec x})
\frac{1}{2}\left(\frac{d{\vec z}_L({\vec q},\eta)}{d\eta}\right)^2  -\delta\tilde\rho_L({\vec z}_L({\vec q},\eta),{\vec x})\,\Phi_L({\vec x},\eta)\right.\\
&&\ \  \qquad \qquad \qquad -  \tilde\rho_L({\vec z}_L({\vec q},\eta),{\vec x}) \left[\frac{1}{2} Q_{\rm TF}^{ij}({\vec q},\eta) \partial_i \partial_j  \Phi_L({\vec x},\eta)+\frac{1}{6} C({\vec q},\eta) \frac{\pd^2}{a^2}  \Phi_L({\vec x},\eta)\right. \nn \\
&&\qquad\qquad\qquad\qquad\qquad\qquad\qquad+\left.\left. \frac{1}{6} Q_{\rm TF}^{ijk}({\vec q},\eta)\partial_i \partial_j\partial_k  \Phi_L({\vec x},\eta)+\cdots \right] \right\}\nn \ ,
\eea
where $(\partial_i \partial_j)_{\rm TF} \equiv \partial_i \partial_j -\tfrac{1}{3}\delta_{ij} \pd^2$, and where we absorbed different coefficients into a redefinition of~$C$. This is to make more simple the comparison with our previous analysis where $C= Q^i_i$.

The action for $\Phi_L$ is the same as in \eqref{EHnewt}. Although not obvious at first, the action starts quadratic in the perturbations. In the unperturbed background, namely when $\vec z = \vec q$, we will have 
\beq
\bar Q^{ij} \propto \delta^{ij} \to \bar Q^{ij}_{\rm TF} = 0,
\eeq
and moreover the term proportional to $C \partial^2\Phi$ becomes a total derivative when its coefficient is $q$-independent. 

Since $\Phi$ is a constrained variable, we can simplify our action by replacing $\Phi$ with the solution to its constraint equation to leading order:
\beq \pd^2\Phi_L = \frac{3}{2} {\cal H}^2 \Omega_m \delta_L + \ldots \ ,
\eeq 
such that the action turns into
\bea
&& S_{\rm LEFT} = \int d \eta \int d^3{\vec x} d^3{\vec q}~ a(\eta)^4  \nn\\
&&\  \left\{\tilde\rho_L({\vec z}_L({\vec q},\eta),{\vec x})  \left[-1-V_S(\vec x,\vec q,\eta)+
\frac{1}{2}\left(\frac{d{\vec z}_L({\vec q},\eta)}{d\eta}\right)^2\right]  - \;\delta\tilde\rho_L({\vec z}_L({\vec q},\eta),{\vec x})\,\Phi_L({\vec x},\eta)\right. \nn\\
&&\quad- \tilde\rho_L({\vec z}_L({\vec q},\eta),{\vec x}) \left[\frac{1}{2} Q_{\rm TF}^{ij}({\vec q},\eta) \partial_i \partial_j  \Phi_L({\vec x},\eta)+\frac{1}{4} C({\vec q},\eta) \cH^2 \Omega_m \delta_L({\vec x},\eta)\right. \nn \\
&&\left.\qquad\qquad \qquad\  \qquad +\left. \frac{1}{6} Q_{\rm TF}^{ijk}({\vec q},\eta)\partial_i \partial_j\partial_k  \Phi_L({\vec x},\eta)+\cdots \right]\right\} \label{effnewt2}\ ,
\eea
up to higher order terms in derivatives and perturbations.  It is straightforward to derive the equations of motion for ${\vec z}_L$, 
\bea
&&\frac{d^2 {\vec z}_L ({\vec q},\eta)}{d\eta^2} + {\cal H} \frac{d {\vec z}_L ({\vec q},\eta)}{d\eta}  =\\
&&\qquad\qquad \vec a_S(\vec q,\eta) - {\vec{\pd}}_x \left[\Phi_L(\vec x,\eta) + \frac{1}{4} \Omega_m \cH^2 C({\vec q},\eta) \delta_L(x)+Q_{\rm TF}^{ij}({\vec q},\eta)\partial_i\partial_j \Phi_L(\vec x,\eta)+ \cdots\right]_{{\vec x=\vec z_L({\vec q},\eta)}}\nn,
\eea
with 
\beq
\vec a_S(\vec q,\eta) \equiv -\left[\vec\partial_x V_S(\vec q,\vec x,\eta)\right]_{{\vec x=\vec z_L({\vec q},\eta)}} \ .
\eeq 
Analogously, for the Poisson equation for the long-distance potential we have 
\beq
\pd_x^2 \Phi_L =  \frac{3}{2} {\cal H}^2\Omega_m  \left[\delta_L({\vec x},\eta)  + \frac{1}{2}\partial_i\partial_j  {\cal Q}_{\rm TF}^{ij}({\vec x},\eta) 
+ \frac{1}{6} \partial_i\partial_j \partial_k  {\cal Q}_{\rm TF}^{ijk}({\vec x},\eta)+\cdots \right]\ ,
\eeq
with
\be
 {\cal Q}^{ij\ldots}_{\rm TF}({\vec x},\eta) = \int d^3{\vec q}\; Q^{ij\ldots}_{\rm TF}({\vec q},\eta)\delta^3({\vec x}-{\vec z}_L({\vec q},\eta)),~{\rm etc}.\
\ee

As expected, we recover the same equations as before. 
For the multipole moments, as well as the potential on short scales, we require a matching procedure: that is they either need to be fitted to observations or be extracted from the UV theory. This last approach is further described in appendix \ref{uvmatch}.

It is interesting to notice that the extra potential term $V_S$, that represents the change in the local energy induced by short distance potentials, as well as the trace of the quadrupole moment, is the main difference with respect to the action in \cite{nrgr1}. Notice also that only the trace-free parts survive in the Poisson equation, whereas scalar components do contribute to the equations of motion~\footnote{This term does not appear for example in binary systems since  $\int R(x) d\tau $ in the action vanishes by the leading order equations or motion, and only a pure counter term remains \cite{nrgr1}. Here, in the continuum limit, these terms do contribute and prove to be essential for the consistency of the theory.}. 


\section{Conclusions}

In this paper we have introduced LEFT: The EFT approach to LSS in Lagrangian space. It is formulated as the theory of a continuum of extended objects that move under the effects of gravitational interactions. Since we are interested in long wavelength correlation functions, we can effectively parametrize the finite size effects with a small number of multipoles. Each of these are in turn characterized by various properties, such as their expectation values, linear response to long wavelength modes, stochastic noise, etc. However to attain a given precision in the final answer only a finite number of coefficients is necessary. 

LEFT is local in space, because there is a hierarchy in space between the non-linear scale and the long-wavelength modes. However, there is no hierarchy in time between the long wavelength modes and the non-linear modes. This implies that, at the level of the equations of motion after replacing the extra terms in LEFT with their response to long-wavelength perturbations, LEFT is non-local in time, as it happens also in the Eulerian EFT~\cite{Carrasco:2013mua,Carroll:2013oxa}. However, both at the level of the equations of motion and of the action, when the extra terms are kept explicit, LEFT is represented as a series of {local} interactions in space and time. Although the non-locality in time of the response functions makes LEFT a peculiar EFT, this has minor consequences at a practical level. Since the Green's function in Fourier space is $k$-independent, in perturbation theory the non-locality in time can be re-absorbed in the redefinition of the time-dependent coefficients~\cite{Carrasco:2013mua}.

We have explicitly performed a one-loop calculation in LEFT for a pure dark matter universe with initial power spectrum characterized by a single power law. We have shown that without the additional terms provided by LEFT, some observable quantities such as the quadrupole of a given Lagrangian region, or the power spectrum of the displacement field, or the power spectrum of the dark matter overdensity, would be divergently large; and even when finite, they would be dependent on the arbitrary scale at which loops are cut off. This is physically unacceptable. We have shown instead that the additional parameters that are needed at a given order in LEFT are sufficient to make these quantities finite, and moreover cutoff independent. At one-loop there are four parameters introduced by LEFT. All of them may be independently determined once enough observable quantities are computed: for example the displacement power spectrum, the expectation value of the quadrupole moment, and the correlation quadrupole-displacement. Of course, if one is interested in only a subset of observables, then one is sensitive only to a particular linear combination of these parameters and so the number of free coefficients that are allowed to be used to match a subset of observations is less than the number of all parameters.

As we have discussed in the introduction, and in sec.~\ref{resum}, there are several expansion parameters that appear in perturbative calculations of dark matter clustering at a given wavelength: the displacement induced by longer wavelength modes $\epsilon_{s<}$, the displacement induced by short wavelength modes $\epsilon_{s>}$, and the curvature induced by longer wavelength modes $\epsilon_{\delta<}$. The previously formulated Eulerian-space EFT of large scale structures has the following disadvantage.  In the Eulerian-space EFT, calculations are performed expanding in all of these parameters. Since in the true universe the parameter  $\epsilon_{s<}$ is not small, calculation have been mainly focussed on IR-safe quantities where the dependence on $\epsilon_{s<}$ cancels out \cite{Carrasco:2013sva,Blas}. However, the reason why the EFT approach is ultimately introduced has to do with the impossibility of describing in a perturbative approach the short distance non-linearities. The fact that the Eulerian approach expands in $\epsilon_{s<}$ is an unfortunate accident of the Eulerian formulation: $\epsilon_{s<}$ does not represent a truly dynamical effect, and it should not therefore affect the convergence of perturbation theory. Contrary to the Eulerian EFT, LEFT does not expand in $\epsilon_{s<}$ at all. In a sense, by going from the Eulerian to the Lagrangian approach, all the $\epsilon_{s<}$ are automatically resummed. The perturbative expansion in LEFT will break down only for those $k$-modes for which the acceleration from the tidal forces of the long modes on a region of the order of the short-distance random displacements is comparable to the acceleration of the center of mass, or when the quadrupole moment of the mass distribution changes significantly the potential, i.e. $\epsilon_{\delta<}(k)\sim 1$ or $\epsilon_{s>}(k)\sim 1$.

LEFT opens up the possibility of a plethora of future directions to explore, such as computing the dark matter power spectrum and correlation functions for our universe, where the effects proportional to $\epsilon_{s<}$ become important, or applying LEFT to compute higher order correlation functions, as well as modeling biased tracers, redshift space distortions  and  considering the fully relativistic version of LEFT which might be important to describe the results of surveys approaching the Hubble volume in size. 
Finally, one should develop the techniques to obtain the parameters of LEFT directly  from multiple different statistics in N-body simulations to check for consistency and to  study their dependence on the particular dark matter models (cold, warm, etc.) and the effects of baryons. We will elaborate on these topics elsewhere.

\subsection*{Acknowledgements}
R.A.P. was supported by NSF grant AST-0807444 and DOE grant DE-FG02-90ER40542, and by the German Science Foundation (DFG) within the Collaborative Research Center (SFB) 676 `Particles, Strings and the Early universe.' R.A.P. would like to thank Imme F. Roewer and Emiliano A. Porto for their patience and support. L. S. is supported by DOE Early Career Award DE-FG02-12ER41854 and by NSF grant PHY-1068380.  M. Z. is supported in part by the National Science Foundation grants PHY- 0855425, AST-0907969, PHY-1213563 and by the David \& Lucile Packard Foundation. We thank Tobias Baldauf, Daniel Baumann, J.J.~Carrasco, Raphael Flauger, Simon Foreman, Daniel Green, Enrico Pajer, and Svetlin Tassev for useful discussions and comments on the draft. 

\appendix

\section{Smoothing}\label{uvmatch}

The basic idea of the EFT is to concentrate on the long distance physics. In this appendix we provide two ways to obtain the coefficients in LEFT from a smoothing procedure of the short distance dynamics. While we find this procedure very instructive and intuitive, we should emphasize that one does not necessarily needs this procedure to construct the long-distance EFT: as we did in the main text, one can just write it directly by using the degrees of freedom and the symmetries that are valid at long distances, and determine the unknown coefficients by fitting to observations. First we will proceed by smoothing the point-particle dynamics at the level of the equations of motion, and later on we will do the same at the level of the action. 

\subsection{Motion of dark matter particles\label{uvmatchone}}

Let us start by taking the full theory of dark matter particles and introduce a window function, $W_{R_0}({\vec q}_1,{\vec q}_2)$, with which we will group the dark matter particles into long-distance cells. The main difference with the analogous construction in the Eulerian EFT approach is that here the window function lives in $\vec q$-space rather than $\vec x$-space, where it would become a time dependent function. Using $W_{R_0}(\vec q_1,\vec q_2)$ we then define the {\it co-moving}
center-of-mass for each cell as:
\beq
 {\vec z}_L ({\vec q}_1,R_0,\eta) \equiv \int d^3{\vec q}_2\; W_{R_0}({\vec q}_1,{\vec q}_2)\; {\vec z}({\vec q}_2,\eta)\ .
 \eeq
The role of $R_0$ here is to provide a cutoff for the correlations which we will be afterwards removed from the theory, i.e. $R_0 \to 0$, while keeping the renormalized (physical) parameter fixed and smoothing-independent. The first entry in $W_{R_0}({\vec q}_1,{\vec q}_2)$, $\vec q_1$, signals the Lagrangian coordinate of the center-of-mass, whereas the second, ${\vec q}_2$, represents all the particles in each given cell. We will adopt a filter that satisfies the following normalization conditions:
\beq 
\int  W_{R_0}({\vec q}_1,{\vec q}_2)\; d^3{\vec q}_2=\int  W_{R_0}({\vec q}_1,{\vec q}_2)\; d^3{\vec q}_1=1\ . 
\eeq 

We now define the short-distance displacement $\delta \vec z(\vec q_1,\vec q_2,\eta)$:
\beq \label{zplusdz}\
 {\vec z}({\vec q}_2,\eta) = {\vec z}_L({\vec q}_1,R_0,\eta) + \delta {\vec z}(\vec q_1,{\vec q}_2,\eta)\ .
\eeq 
which satisfies
\beq
\int d^3{\vec q}_2\; W_{R_0}({\vec q},{\vec q}_2)({\vec z}({\vec q}_2,\eta) - {\vec z}_L({\vec q},R_0,\eta)) = 0\ .
\eeq
The short displacement $\delta \vec z(\vec q_1,\vec q_2,\eta)$ depends on two variables, same as our window function, where $\vec q_1$  accounts for the label of each cell, whose center-of-mass is represented by $\vec z_L(\vec q_1,R_0,\eta)$, and $\vec q_2$ which serves as the Lagrangian position of each particle on a given cell.

Let us now start with the Poisson equation.  Consider the mass density,
\beq
\label{fulldelta}
1+\delta_m({\vec x},\eta) = \int d^3 {\vec q}_2~\delta^3( {\vec x} - {\vec z}({\vec q}_2,\eta))\ ,
\eeq
which we can multiply by $\int W_{R_0} ({\vec q}_1, {\vec q}_2)d^3{\vec q}_1=1$ to obtain
\beq
1+\delta_m({\vec x},\eta) = \int d^3{\vec q}_1\; d^3 {\vec q}_2~W_{R_0} ({\vec q}_1,{\vec q}_2)~\delta^3( {\vec x} - {\vec z}({\vec q}_2,\eta))\ .
\eeq
We now expand this expression in a Taylor series around $\vec z_L(\vec q_1,R_0,\eta)$ to obtain:
\be
\delta_{m,L}({\vec x},\eta) = \delta_{n,L}({\vec x},\eta)  + \frac{1}{2} \partial_i\partial_j  {\cal Q}^{ij}_{R_0}({\vec x},\eta) - \frac{1}{6} \partial_i\partial_j \partial_k {\cal Q}_{R_0}^{ijk}({\vec x},\eta) + \ldots\ ,
\eeq
where we have introduced:
\bea
\label{eqsource}
1+\delta_{n,L}({\vec x},\eta)&\equiv& \int  d^3{\vec q}~\delta^3( {\vec x} - {\vec z}_L({\vec q},\eta))\ , \nonumber \\
\label{calqn}  {\cal Q}_{R_0}^{i \dots i_p}({\vec x},R_0,\eta) &\equiv& \int d^3 {\vec q}~ Q^{i\ldots i_p}_{R_0}({\vec q},R_0,\eta)\delta^3({\vec x}-{\vec z}_L({\vec q},\eta))\ ,
\eea
with
\beq
\label{qij0}
 Q^{i\ldots i_p}_{R_0}({\vec q}_1,\eta) \equiv \int d^3{\vec q}_2\;  W_{R_0}({\vec q}_1,{\vec q}_2)\; \delta z^i({\vec q}_2,{\vec q}_1,\eta)\ldots \delta z^{i_p}({\vec q}_2,{\vec q}_1,\eta)\ ,
\eeq
being the multipole moments, which depend on $R_0$ through the smoothing function and on the short dynamics through the displacements. 
The first line in \eqref{eqsource} gives the over density of the centers of mass and the second the corresponding multipole moments.

We can now define a long wavelength potential $\Phi_L$ as the one sourced by the density field obtained using the above expansion:
\beq
\label{phiL1}
\partial^2_x \Phi_L \equiv \frac{3}{2} {\cal H}^2 \Omega_m \delta_{m,L}({\vec x},\eta) = \frac{3}{2} {\cal H}^2 \Omega_m\left(\delta_{n,L}({\vec x},\eta)  + \frac{1}{2} \partial_i\partial_j  {\cal Q}^{ij}_{R_0}({\vec x},\eta) - \frac{1}{6} \partial_i\partial_j \partial_k {\cal Q}_{R_0}^{ijk}({\vec x},\eta) + \cdots\right) \ .
\eeq
The full potential is given by this long-wavelength term, plus a contribution from short modes, i.e.   $\Phi = \Phi_L+ \Phi_S$, which satisfies:
\beq\label{phi_tot}
\partial^2_x \Phi = \frac{3}{2} {\cal H}^2 \Omega_m \delta_m({\vec x},\eta)\ ,
\eeq
with the full matter density.

To obtain the equation of motion for the centers of mass, we convolve the Lagrangian equations of motion with the window function,
\beq
\frac{d^2 {\vec z}({\vec q},\eta)}{d\eta^2} + {\cal H} \frac{d {\vec z}({\vec q},\eta)}{d\eta}  =  - \vec\partial_x \Phi[{\vec z}({\vec q},\eta)]\ ,
\eeq
and get
\beq
\frac{d^2 {\vec z}_L ({\vec q}_1,R_0,\eta)}{d\eta^2} + {\cal H} \frac{d {\vec z}_L ({\vec q}_1,R_0,\eta)}{d\eta}  =  - \int d^3{\vec q}_2\; W_{R_0}({\vec q}_1,{\vec q}_2)\;  \vec\partial_x \Phi[{\vec z}({\vec q}_2,\eta)]\ .
\eeq
We next use the splitting $\Phi = \Phi_L+ \Phi_S$, and as we did before expand $\Phi_L$ as a long wavelength quantity in Taylor series around ${\vec z}_L(\vec q_1,R_0,\eta)$:
\bea
&& \int d^3{\vec q}_2\; W_{R_0}({\vec q}_1,{\vec q}_2)\; \Phi_L[{\vec z}({\vec q}_2,\eta)] = \\ 
&=& \Phi_L[{\vec z}_L({\vec q}_1,R_0,\eta)]  +\frac{1}{2} \partial_i\partial_j \Phi_L[{\vec z}_L({\vec q}_1,R_0,\eta)] \int d^3{\vec q}_2\; W_{R_0}({\vec q}_1,{\vec q}_2)\; \delta {\vec z}^i({\vec q}_2,{\vec q}_1,\eta)\delta {\vec z}^j({\vec q}_2,{\vec q}_1,\eta) + \cdots \nn\ ,
\eea
The equation of motion thus becomes
\bea
\frac{d^2 {\vec z}_L ({\vec q},R_0,\eta)}{d\eta^2} + {\cal H} \frac{d {\vec z}_L ({\vec q},R_0,\eta)}{d\eta}  &=& \vec a_S(\vec q,\eta) - \vec\partial_x \left[\Phi_L(\vec x,\eta) + \frac{1}{2}  Q^{ij}_{R_0}({\vec q},\eta)\partial_i\partial_j \Phi_L({\vec x},\eta) \right.\nn\\ 
&+&\left. \frac{1}{6}  Q^{ijk}_{R_0}({\vec q},\eta)\partial_i\partial_j \partial_k \Phi_L(\vec x,\eta)+ \cdots\right]_{\vec x = {\vec z}_L({\vec q},R_0,\eta)}\ ,
\eea
where we have defined:
\be
\vec a_S(\vec q,\eta)=-\int d^3{\vec q}_2\;  W_{R_0}({\vec q}_1,{\vec q}_2)\; \vec \d_x \Phi[\vec z(\vec q_2,\eta)] \equiv - \left[\vec\partial_x  V_S(\vec q,\vec x,\eta)\right]_{\vec x = {\vec z}_L({\vec q},R_0,\eta)}\ ,
\ee
with 
\beq
V_S(\vec q_1,\vec z_L(\vec q_1,\eta),\eta) =\int_k  e^{i \vec k \cdot \vec z_L(\vec q_1,\eta)} \Phi_S(\vec k,\eta) \int d^3{\vec q}_2\; W_{R_0}({\vec q}_1,{\vec q}_2) \;e^{i \vec k \cdot \delta \vec z(\vec q_1,\vec q_2,\eta)}\ .
\eeq

Both the multipole moments and $V_S$ depend on the short dynamics and thus cannot be calculated directly within the theory. The relation between their values and the EFT variables necessarily involves free parameters. In the main text we divided these fields into an expectation value, a stochastic term and a response. We wrote the most general expression for these fields in a derivative expansion of the long wavelength fields. At any given order in perturbation theory this procedure introduces a finite number of free parameters that has to be matched with computations using the UV theory (for example exactly solved by using numerical simulations) or fitted to data.  From the previous analysis, however, we learned that the new term is the gradient of a scalar, and therefore,
\beq
V_S(\vec q,\vec z_L(\vec q,\eta),\eta) = \frac{3}{2} \Omega_m \cH^2  l_{\Phi_S}^2(\eta) \vec \partial_q \cdot \vec z_L(\vec q,\eta) + \ldots\ ,
\eeq
which precludes a term of the sort $\partial_q^2 \vec z_L(\vec q,\eta)$ in $\vec a_S$.

As written in equation \eqref{phiL1} $Q^{ii}$ sources $\Phi_L$, which in turn produces a force on the center of mass. This force clearly has the same form as the one from $\vec a_S({\vec q},\eta)$, and thus the split between $\Phi_L$ and $\vec a_S({\vec q},\eta)$ is a matter of definition (as long as we deal with correlations of the displacement only). We could equivalently have defined a different smoothed potential $\tilde \Phi_L$ satisfying:
\beq
\partial^2_x \tilde \Phi_L = \frac{3}{2} {\cal H}^2 \Omega_m\left(\delta_{n,L}({\vec x},\eta)  + \frac{1}{2} \partial_i\partial_j  {\cal Q}^{ij}_{{\rm TF},R_0}({\vec x},\eta) + \cdots\right).
\eeq
The different definition results in a change in the value of $l_{\Phi_s}^2(\eta)$. The $\tilde \Phi_L$ potential is perhaps more standard because it is only sourced by the trace free part of the quadrupole moment, which is the only relevant quality to compute the forces outside the particle. In fact for multipole moment $l$ there are $(2l+1)$  coefficients needed to describe the potential outside the mass distribution. 

In addition to correlation functions of the displacement we are interested in computing correlation functions of the density field, thus we now examine more carefully the expression for the density field on large scales. Using equation (\ref{fulldelta}) we compute the Fourier components of the density:
\be
\begin{split}
 \delta_m(\vec k,\eta) &= \int d^3 q_2 e^{- i \vec k \cdot {\vec z}({\vec q}_2,\eta)} \\
&= \int d^3 q_1 d^3 q_2 W_{R_0} ({\vec q}_1, {\vec q}_2) e^{- i \vec k \cdot {\vec z}({\vec q}_2,\eta)} \\
&=  \int d^3 q_1  e^{- i \vec k \cdot {\vec z}_L({\vec q}_1,\eta)} \int d^3 q_2 W_{R_0} ({\vec q}_1, {\vec q}_2) e^{- i \vec k \cdot \delta{\vec z}({\vec q}_1,{\vec q}_2,\eta)} \\
&=  \int d^3 q_1  e^{- i \vec k \cdot {\vec z}_L({\vec q}_1,R_0,\eta)} \exp\left[ \sum_n \frac{(-i)^n}{n ! } k_{i_1} \cdots k_{i_n} Q^{i_1 \cdots i_n}_{R_0,c}({\vec q}_1,\eta)\right]
\end{split}
\ee
where $Q^{i_1 \cdots i_n}_{R_0,c}$ are the connected version of the $N$-point multipoles. Clearly if we are interested in the density for small enough $\vec k$, small compared to inverse of the displacements $ \delta{\vec z}({\vec q}_1,{\vec q}_2,\eta)$, this expression can be expanded in a Taylor series and the resulting expression is nothing other than the Fourier transform of what we called $\delta_{L}({\vec x},\eta)$ in equation (\ref{phiL1}). 
Physically this Taylor expansion is very reasonable as the typical displacements when we smooth over small region comparable or smaller than the non-linear scale $k^{-1}_{NL}$, is also comparable to the non-linear scale. In other words the terms in the Taylor expansion are suppressed by powers of $k$ times the typical relative displacements between particles that started in a small region of size $R_0$. So the expansion is basically an expansion in $k/k_{\rm NL}$.

\subsection{The coefficients in the effective action}\label{matching}

 To obtain the multipole moments in the effective action we can compute the displacement and extract the coefficients from N-body simulations, or we can match directly with the $UV$ theory, namely with our original action in \eqref{action2}, that is indeed valid to all scales.
In order to find the relationship between the quantities computed in the two theories,  we can use the same smoothing procedure we discussed previously, now at the level of the action. We start by multiplying our original action $S_{\rm pp}^{\rm DM}$ in \eqref{actiondm} by $\int d^3{\vec q}_2\; W_{R_0}({\vec q},{\vec q}_2) = 1$. Moreover we split $\Phi$ as $\Phi=\Phi_S+\Phi_L$, where $\Phi_S$ and $\Phi_L$ are defined as in \eeqref{phiL1} } and (\ref{phi_tot}): a short  potential on scales of order $R_0$ and below, and longer distance modes, respectively. Then we proceed by splitting the action as 
\bea
&&S_{\rm pp}^{\rm DM}=\int a^4 d\eta \int d^3{\vec x} \int d^3{\vec q} d^3{\vec q}_2~W_{R_0}({\vec q},{\vec q}_2) \ \times\\ \nn
&& \qquad\qquad\left\{\tilde\rho({\vec z}({\vec q}_2,\eta),{\vec x})\left[-1+\frac{1}{2}\left(\frac{d{\vec z}({\vec q}_2,\eta)}{d\eta}\right)^2  \right]-\delta\tilde\rho({\vec z}({\vec q}_2,\eta),{\vec x}) (\Phi_S({\vec x},\eta)+\Phi_L({\vec x},\eta))\right\}\ .\nn  \label{actionW}
\eea 

We now follow a similar procedure to appendix \ref{uvmatchone}, and write the position of each particle as a {\it bulk} motion, representing the movement of the center-of-mass of a collection of dark matter particles within the region $R_0$, plus a short-distance displacement relative to this center, namely 
\be
\label{expz2}
{\vec z} ({\vec q}_2,\eta) = {\vec z}_L({\vec q},\eta) + \delta {\vec z}({\vec q},{\vec q}_2,\eta)\ , 
\eeq
with
\be\label{eq:zshort}
\int d^3 q_2\; W_{R_0}(\vec q_1,\vec q_2)\; \delta{\vec z}({\vec q},{\vec q}_2,\eta)=0\ .
\ee

The action now admits a multipole expansion for the long distance modes, as we discussed already in sec. \ref{sec:exp}. In particular expanding \eqref{actionW} in powers of $\delta {\vec z}$ will turn \eeqref{action2} into an expression as in~\eqref{effnewt}, from which we can read off the form of our multipole moments and the coefficients associated
with $V_S$. Let us do this in steps. First, let us concentrate on the $\Phi_L$ part, then using
\beq
\delta\tilde\rho_L({\vec z}({\vec q}_2,\eta),{\vec x}) = \delta\tilde\rho_L ({\vec z}_L({\vec q},\eta),{\vec x}) + \partial_i \tilde\rho_L ({\vec z}_L({\vec q},\eta),{\vec x})\delta { z}^i  + \frac{1}{2} \partial_i\partial_j \tilde\rho_L ({\vec z}_L({\vec q},\eta),{\vec x}) \delta { z}^i\delta { z}^j + \cdots,
\eeq
and plugging it back into the $\Phi_L$-action, integrating by parts, we obtain
\bea
&&S_{\rm pp}^{\rm DM}\supset\\ \nn
&&-\int a^4 d\eta \int d^3{\vec x} d^3{\vec q}\left\{\ldots + \tilde\rho_L ({\vec z}_L({\vec q},\eta),{\vec x}) \left(\int d^3{\vec q}_2\;W_{R_0}({\vec q},{\vec q}_2) \delta {\vec z}^i\delta {\vec z}^j\right) \frac{1}{2}\partial_i \partial_j \Phi_L ({\vec x},\eta) + \cdots \right\},
\eeqa
 after taking into account of \eeqref{eq:zshort}.
From here we can identify the expression for the multipoles:
\bea
\label{matchqij} \nn
&&Q^{i_1i_2\cdots i_n}_{\rm TF}({\vec q},\eta) = \left(\int d^3{\vec q}_2\; W_{R_0}({\vec q},{\vec q}_2)\; \delta { z}^{i_1}({\vec q},{\vec q}_2,\eta)\delta { z}^{i_2}({\vec q},{\vec q}_2,\eta)\ldots \delta { z}^{i_n}({\vec q},{\vec q}_2,\eta)\right)_{\rm TF}\ ,\\
&& C({\vec q},\eta) =  Q^i_i(\vec q,\eta)= \left(\int d^3{\vec q}_2 ~W_{R_0}({\vec q},{\vec q}_2)\; \delta { z}^i({\vec q},{\vec q}_2,\eta)\delta { z}^i({\vec q},{\vec q}_2,\eta) \right)\ ,
\eea
and so on and so forth to all order in derivatives.

For the other terms, first of all it is straightforward to see that the sum of kinetic terms will split into the motion of the
center of mass plus the internal kinetic energy $K(\vec q,\eta)$, namely
\bea\label{kinetic}
&& S_{\rm pp}^{\rm DM}\supset \frac{1}{2}\int d\eta d^3 \vec x d^3 \vec q~\bar \rho_m(\eta) a^4 \delta(\vec x - \vec z(\vec q,\eta))\left(\frac{d \vec z(\vec q,\eta)}{d\eta}\right)^2\\ \nn
&& \quad\quad
\to\quad 
\int d\eta d^3 \vec x d^3 \vec q~\bar \rho_m(\eta) a^4 \delta(\vec x - \vec z_L(\vec q,\eta)) \left[\frac{1}{2}\left(\frac{d \vec z_L(\vec q,\eta)}{d\eta}\right)^2 + K(\vec q,\eta)\right]\ ,  
\eea
with
\beq
K(\vec q_1,\eta) = \frac{1}{2}\int d^3 \vec q_2 W_{R_0}(\vec q_1,\vec q_2) \delta \dot z^i(\vec q_1,\vec q_2,\eta) \delta \dot z^i(\vec q_1,\vec q_2,\eta)\ .
\eeq
After doing the integral in $d^3 \vec x$ in equation (\ref{kinetic}) one can see that the term containing $K(\vec q,\eta)$ is independent of the long wavelength fields. It can thus be ignored. 

On the other hand for the short mode of the potential, we have
\beq
S_{\rm pp}^{\rm DM}\supset\int d^3 \vec x d^3 \vec q_1 d^3 \vec q_2\; W_{R_0}(\vec q_1,\vec q_2)\; \tilde\rho({\vec z}({\vec q}_2,\eta),{\vec x})\, \Phi_S({\vec x},\eta)\ ,
\eeq
which leads to the term
\bea
&&S_{V_S}\equiv \int d^3 x d^3 q_1 d^3 q_2\; W_{R_0}(\vec q_1,\vec q_2)\;\bar\rho_m(\eta)\, \delta^3(\vec x -\vec z(q_2,\eta))\; \Phi_S(\vec x,\eta) \\ \nonumber
&&\qquad =\int d^3 q_1 d^3 q_2\;  W_{R_0}(\vec q_1,\vec q_2)\;\bar\rho_m(\eta)\, \int_k~\Phi_S(\vec k,\eta)~e^{i\vec k\cdot [\vec z_L(\vec q_1,\eta) + \delta \vec z(\vec q_1, \vec q_2,\eta)]} \ .
\eea
Hence we write
\beq
S_{V_S} = \int d^3 x d^3 q \;\bar\rho_m(\eta)\,\delta^3 (\vec x - \vec z_L(\vec q,\eta))\; V_S(\vec q, \vec x,\eta) \ ,
\eeq
with
\beq
V_S(\vec q_1,\vec k,\eta) =\int d^3x\; e^{- i\vec k\cdot\vec x} \;V_{S}(\vec q_1,\vec x,\eta)\equiv \int d^3 q_2 \; W_{R_0}(\vec q_1,\vec q_2)\; \Phi_S(\vec k,\eta)\; e^{i\vec k\cdot \delta \vec z(\vec q_1,\vec q_2,\eta)}\ .
\eeq
As for the case of the multipoles, this potential must be split into a noise term plus response. The reader will notice these are the same expressions we obtained previously.

\section{Dimensional Regularization}\label{app:dimreg}

Here we discuss the case of a power law universe $P_L = A(\eta_0) k^n$, using dimensional regularization where we analytically continue
the number of space dimensions to $d = 3-\epsilon$. Even though our universe does not have a power law power spectrum at late times, it is still instructive to analyze the divergences for power law universes. Moreover, within a range of momenta near the non-linear scale, the power spectrum does resemble a power law universe with $n\simeq -2$, and therefore the intuition built from power law universes can be useful.

We noted before that the scaling of each term in LEFT depends on $n$. As it was discussed in \cite{Carrasco:2012cv, Pajer:2013jj}, this is related to the UV divergences in the one-loop computations, that occur for $n \geq -1$ when we remove the cutoff of the theory. Notice that the Einstein-de Sitter universes with a single power law power spectrum are endowed with a scaling symmetry. To ensure that this symmetry is not broken by the regularization procedure of the loop integrals, one can regularize using dimensional regularization, as we demonstrate in what follows. There is one subtlety worth emphasizing. In dimensional regularization one introduces a new dimensionful parameter, traditionally denoted as $\mu$, which one can interpret as the scale at which the theory is renormalized. Now the answer will depend not only on $k/k_{\rm NL}$, but also on $k/\mu$. However, since in dimensional regularization scaleless integrals are set to zero (i.e. power-law divergences are not present, unlike cutoff regularized integrals), only logarithmic divergences appear. Therefore the general answer will only introduce extra factors of $\log (k/\mu)$. As usual, $\mu$-independence leads to a renormalization group flow for the couplings of the theory.

In practice, we define the $d$-dimension power spectrum as~\footnote{The implementation of dimensional regularization we perform here breaks diffeomorphism invariance, already at zeroth order. Notice we did not specified the gauge of the power spectrum in~\eeqref{eq:power_dim_reg}. This is not a problem for us in this context, since are working at Newtonian level. See for example~\cite{Senatore:2009cf} for a detailed discussion on the context of Inflation.}
\be\label{eq:power_dim_reg}
P^{(\epsilon)}_L(k)=\frac{k_\mu{}^\epsilon}{k_{\rm NL}^3}\left(\frac{k}{k_{\rm NL}}\right)^n\ ,\quad
\ee
where the comoving wavenumber $k_\mu$ has been inserted to keep $\delta(\vec x)$ in real space dimensionless:
\be
\langle\delta(\vec x)^2\rangle=\int \frac{d^{3-\epsilon}k}{(2\pi)^{3-\epsilon}} P_L^{(\epsilon)}(k)\ .
\ee
We could have equivalently defined a physical time-dependent renormalization scale $\mu(\eta)= k_\mu a(\eta)$, in which case we would have replaced $k_\mu$ with $\mu\, a(\eta)$. 
The loop integrals can be readily  evaluated, and the resulting correlation functions for the displacement read: 
\beq
\!\!\!\!C^{(22)}_{ij}(k,\eta) = \frac{9}{49} \pi^{\tfrac{-\epsilon}{2}-1}   A^2(0) a(\eta)^{4(1-\epsilon)}    \frac{(2-\epsilon)(4-\epsilon)\Gamma\left[-n+\tfrac{\epsilon}{2}+\tfrac{1}{2}\right]\Gamma\left[\tfrac{1}{2}(n+3-\epsilon)\right]}{2^{(n+9-\epsilon)} \Gamma\left[2-\tfrac{n}{2}\right]^2\Gamma\left[\tfrac{1}{2}(n+4-\epsilon)\right]}~\frac{ k^i  k^j}{k^{1-2 n+\epsilon}} k_\mu^{2\epsilon} \ ,
\eeq
and
\beq
\!\!\!\!C^{(13)}_{ij}(k,\eta) = \frac{5}{672}  \pi^{\tfrac{-\epsilon-1}{2}}  A^2(0) a(\eta)^{4(1- \epsilon)} \frac{(2-\epsilon)(4-\epsilon)\Gamma\left[\tfrac{-\epsilon+5}{2}\right] {\rm Csc}\left(\tfrac{1}{2} \pi(n-\epsilon-1)\right)}{\Gamma\left[2-\tfrac{n}{2}\right]\Gamma\left[\tfrac{n}{2}-\epsilon+4\right]} ~\frac{ k^i  k^j}{k^{1-2 n+\epsilon}} k_\mu^{2\epsilon}\label{c13dr}\ ,
\eeq
where we used $D(\eta) \propto a^{d-2}(\eta)$, and ${\rm Csc}[z]$ stands for the {\it Cosecant}. Note that the IR divergences cancel out individually in each term.

To illustrate how to proceed, let us take a power law universe with $n=-1$, where logarithmic divergences appear in $C^{(13)}_{ij}(k,\eta)$, while $C^{(22)}_{ij}(k,\eta)$ remains finite. Evaluating the expression in \eeqref{c13dr} for $n=-1$, and using the standard expansion of the $\Gamma[z]$ function near its poles, we have
\beq
\label{pole}
 C_{ij}^{(13)}(k,\eta) = \frac{4}{63\pi^2} a^4(\eta) A^2(0) \left(\frac{1}{\epsilon} -\log k+2\log k_\mu\right) \frac{ k^i  k^j}{k^3} + \ldots\ ,
\eeq
in the limit $\epsilon \to 0$.

As we discussed in sec.~\ref{sec:comb}, the equation for the displacement in LEFT takes the form (see~\eeqref{thetaeq}):
\beq
\ddot {\vec s}(\vec q,\eta) + \cH\dot {\vec s}(\vec q,\eta) -\frac{3}{2}\cH^2\Omega_m \vec s(\vec q,\eta) = \frac{3}{2}\cH^2\Omega_m l^2_{s,\rm comb}\partial_q^2 \vec s(\vec q,\eta) +\ldots,
\eeq
with the combined parameter $l^2_{s, \rm comb}$ defined in \eeqref{eq:lscomb}, having units of a comoving length. According to \eeqref{pole}, the divergence in the correlation of the displacement is given by
\beq
C_{ij}^{(13)}(k,\eta)+C_{ij}^{(31)}(k,\eta)\ \ \toepsilon\  \ \frac{8}{63\pi^2} A^2(\eta)\frac{1}{\epsilon}~\frac{ k^i  k^j}{k^3} + \ldots\  ,
\eeq
with $A(\eta) = A(\eta_0)a^2(\eta)$. Following similar steps as in sec. \ref{sec:comb}, the counter-term becomes (for an Einstein de Sitter universe) 
\beq
\vec s^{~\rm c.t.}(k,\eta) =\frac{1}{6} (l^{\rm c.t.}_{s, \rm comb})^2~\partial_q^2 \vec s(k,\eta)\ ,
\eeq
hence
\bea
&&\langle s_i^{(1)}(k,\eta) s_j^{\rm c.t.}(k,\eta)\rangle =\\ \nn
&& \qquad\quad -\frac{k^2}{6} (l^{\rm c.t.}_{s, \rm comb})^2 C_{ij}^{(11)}(k,\eta) = -\frac{k^2}{6}  (l^{\rm c.t.}_{s, \rm comb})^2 \frac{k^ik^j}{k^4} P_L^{(\epsilon)}(k,\eta) = -\frac{k^2}{6} A(\eta) (l^{\rm c.t.}_{s, \rm comb})^2 \frac{k^ik^j}{k^3} k_\mu^\epsilon\ .
\eea
Therefore, combining both possible contractions, and choosing (recall, for $n=-1$, $k^{-2}_{\rm NL} \equiv A/2\pi^2$)
\beq
(l^{\rm c.t.}_{s, \rm comb})^2 = A(\eta)\frac{24}{63\pi^2}\frac{1}{\epsilon} = \frac{48}{63} \frac{1}{k_{\rm NL}^2}\frac{1}{\epsilon}\ ,
\eeq
the final expression for the correlation of the displacement to one-loop order becomes
\beq
(k^3 k^i  k^j) C^{\rm 1loop}_{ij}(k,\eta) = \frac{k^2}{k_{\rm NL}^2} \left\{1 +  \frac{k^2}{k_{\rm NL}^2} \left( {\rm finite} - \frac{1}{6} \left(l^{\rm ren}_{s, \rm comb}(k_\mu,\eta)\right)^2\; k_{\rm NL}^2 -\frac{4}{63} \log (k/k_\mu)\right)\right\}\label{cij}\ ,
\eeq
where we introduced the renormalized parameter $l^{\rm ren}_{s,\rm comb}(k_\mu,\eta)$ that requires a matching procedure, as we explained in sec. \ref{sec:reno}. The $k_\mu$ dependence in this coefficient must cancel out against a similar factor in the logarithmic term, which leads to a (somewhat trivial) renormalization group equation: 
\beq
k_\mu \frac{d}{dk_\mu} (l^{\rm ren}_{\rm comb})^2 (k_\mu,\eta) =  -\frac{24}{63} k_{\rm NL}^{-2}\ .
\eeq
The procedure follows similar steps for (logarithmic) divergences at any value of $n$ as previously discussed. For power law divergences, renormalized parameters are still required, however, counter-terms may not be needed. In dimensional regularization the power counting is therefore straightforward, without the necessity of accounting for an extra scale, a cutoff $\Lambda$, in the problem.

In dimensional regularization it is thus quite simple to see from the scaling of our multipole moments when (logarithmic) divergences might arise. Since these will absorb the divergent parts of the loop integrals, they need to scale precisely in the right fashion to absorb the $1/\epsilon$ poles that appear. It is then easy to show that divergences will occur for $n=-1+2p$, with~$p \geq 1$.

\section{Simple Toy model}\label{toymodel}

In section \ref{resum}, we  argued that for some range of initial conditions, which include those applicable to our universe, it would be advantageous to resum higher order terms in the displacement. At lowest order in the density contrast, this corresponds to considering and expression like 
\be
\delta(\vec x)=\delta^{(1)}+d_k^{(1)} \partial_k \delta^{(1)}+ {1\over 2} d_k^{(1)} d_l^{(1)} \partial_k\partial_l\delta^{(1)} + \cdots = \delta^{(1)}(\vec x +  \vec{d} )\ .
\ee
Although discussing this in detail is beyond the scope of this paper, we want to illustrate the consequences of such an approximation using a simple toy model.

We  will consider a case, analog to CMB lensing, in which the density field is a Gaussian random field $\delta_L$ which is shifted by a displacement field $\psi$ which is also a Gaussian. The field $\delta_L$ has power spectrum $P_L(k)$ and the field $\psi$ has power spectrum $P_\psi(k)$. For simplicity, we will work in one spatial dimension and take $\delta$ and $\psi$ as uncorrelated. The model is then
\bea
\delta(x)&=&\delta_L(x+\psi(x)) \nonumber \\
&=& \int {dk \over 2\pi} \hat\delta_L(k) e^{- i k (x + \psi(x))}\ .
\eea

In this toy model the correlation function can be computed exactly
\bea\label{zeta1} 
\xi(x)&=&\langle \delta(x)\delta(0)\rangle = \int {dk \over 2\pi} P_L(k) e^{i k x} \langle e^{i k (\psi(x)-\psi(0))} \rangle \nonumber \\
&=& \int {dk \over 2\pi} P_L(k) e^{i k x}  e^{- k^2 \Delta_\psi(x)/2}\ ,
\eea
where we have define $\Delta_\psi(x)= \langle(\psi(x)-\psi(0))^2\rangle$. Note that 
\bea
\Delta_\psi(x)&=& 2(\langle(\psi(0))^2\rangle - \langle\psi(x)\psi(0)\rangle) \nonumber \\
&=&  \int {dk \over 2\pi} P_\psi(k) (1-e^{i k x})\ ,
\eea
thus modes with $k x \ll 1$ do not contribute. They shift both points in the correlation function by the same amount.

An important point is that, in equation (\ref{zeta1}),  the contribution to the correlation function  at a separation $x$ coming from modes of wavenumber $k$ is suppressed by $\Delta_\psi$, which receives contributions from all the modes with momentum larger than $1/x$, even those that have momentum smaller than $k$. In other words, in order to contribute, the $\psi$-modes only need to be UV with respect to $1/x$, not with respect to $k$. 

In this toy model, the one-loop answer corresponds to expanding the exponential in equation (\ref{zeta1}) to first order in $P_\psi$. We get
\bea\label{zeta1loop} 
\xi^{\rm 1-loop}(x)&=& - \int {dk \over 2\pi} P_L(k) e^{i k x}  k^2 \frac{\Delta_\psi(x)}{2} \nonumber \\
&=& \langle(\psi(0))^2\rangle \xi''(x) - \langle\psi(x)\psi(0)\rangle \xi''(x) \nonumber \\
&\equiv& \xi_{13}(x) + \xi_{22}(x)\ ,
\eea
where we have called the two contributions $\xi_{13}$ and  $\xi_{22}$ because they come from what in the power spectrum we would usually call the $13$ and $22$ terms. Again, notice that the IR cancelation is only for modes that are IR with respect to $1/x$. This runs contrary to the standard intuition:  that is that in the power spectrum at mode $k$ the IR cancellation happens for modes that are long compared to $k$ and not $1/x$.

The one-loop power spectrum is just the Fourier transform of the one-loop correlation function. It is given by:
\be
P^{\rm 1-loop}(k) = P_{13} + P_{22} = -  \langle(\psi(0))^2\rangle k^2 P_L(k) + \int {dk^\prime \over 2 \pi} P_\psi(k^\prime) (k-k^\prime)^2 P_L(k-k^\prime)\ . 
\ee
This can be written in a suggestive way:
 \be
P^{\rm 1-loop}(k) =  \int {dk^\prime \over 2 \pi} P_\psi(k^\prime) [(k-k^\prime)^2 P_L(k-k^\prime)- k^2  P_L(k)]\ . 
\ee
This expression clearly shows the cancelation in the limit $k^\prime \ll k$.  This is the source of the standard  intuition that modes with  $k^\prime \ll k$ are not relevant for the power spectrum at wavenumber $k$, for example for long modes not being relevant for the damping of the BAO peaks in Fourier space.

One may wonder how does this relate to the previous claim based on inspection of the formulas for the correlation function. We can explicitly write
\bea
\xi^{\rm 1-loop}(x)&=& \int {dk \over 2 \pi} e^{i k x} P^{\rm 1-loop}(k) \nonumber \\
&=& \int {dk \over 2 \pi} {dk^\prime \over 2 \pi} e^{i k x} P_\psi(k^\prime) [(k-k^\prime)^2 P_L(k-k^\prime)- k^2  P_L(k)] \nonumber \\
&=& \int {dk \over 2 \pi} {dk^\prime \over 2 \pi} e^{i k x} P_\psi(k^\prime) k^2  P_L(k) [e^{i k^\prime x} - 1]\ .
\eea
One has to be very careful because when computing the correlation function one is integrating over both $k$ and $k^\prime$ and the exponential is oscillating so much. The same pair of $k$-$k^\prime$ appears in both integrals but with a different phase and the cancellation only happens for $k^\prime x \ll1$. So for a fixed $k$ in the power spectrum there is no contribution from $k^\prime\ll k$ but when we go to the correlation function and integrate over both $k$ and $k^\prime$ the cancellation between the $22$ and $13$ contributions only happens for $k^\prime x \ll 1 $ regardless of $k$.

This simple expressions for the correlation function allow us to estimate the corresponding size of the corrections in perturbation theory.
The correction we have computed in this toy model, that only includes the effects of the displacements and not the dynamical effects, is
\bea
\Delta\xi^{\rm 1-loop}(x) &=& {1 \over 2 } \xi''(x) \Delta_\psi(x)\nonumber \\
&=& {1 \over 2 }  x^2 \xi''(x) {\Delta_\psi(x) \over x^2}\ .
\eea 
In our universe, this term is very enhanced, basically because of the sharpness of the BAO peak which around that scale gives $x^2 \xi''(x)/\xi \sim 150$ and ${\Delta_\psi(x) / x^2 \xi(x)}\sim 3$. This means that this correction is much larger than they dynamical correction which is of order $\xi(x)^2$.  
\be
\Delta\xi^{\rm 1-loop}(x)\sim  {1 \over 2 } \xi''(x) \Delta_\psi(x) {1 \over 2 } \sim \xi(x) \gg \xi(x)^2,
\ee
at the BAO scale. Another important thing to keep in mind is that, as the toy model above illustrates, the effect of $\Delta_\psi(x)$ is to reduce the contribution of the high $k$ modes to the correlation function, thus broadening the peak. But, as is well known, most of this effect will be substantially reduced by the so-called reconstruction procedure that tries to undo precisely these motions.

\end{document}